\RequirePackage{fix-cm}
\RequirePackage{tikz}
\documentclass[twocolumn]{svjour3}          
\smartqed  
\usepackage{ifpdf}

\usepackage[utf8]{inputenc} 
\usepackage{graphicx}
\usepackage{epsfig}
\usepackage{subfigure}
\usepackage{calc}
\usepackage{amssymb}
\usepackage{amstext}
\usepackage{amsmath}

\usepackage{amsthm}
\usepackage{multicol}
\usepackage{pslatex}
\usepackage{longtable}
\usepackage{caption,multicol,booktabs,array} 
\usepackage[numbers]{natbib}
\usepackage{microtype}
\usepackage{verbatim} 
\usepackage{comment}

\usepackage{hyperref}
\usepackage[utf8]{inputenc}
\usepackage[euler]{textgreek}
\usepackage[draft]{minted}
\usepackage{listings,xcolor}
\usepackage{multicol}
\usepackage{graphicx}
\usepackage{float}
\usepackage{graphics}
\usepackage{pdflscape}
\usepackage{multirow}
\usepackage{makecell}
\usepackage{adjustbox}
\usepackage{rotating}
\usepackage{paralist} 
\usepackage{longtable} 
\usepackage{comment} 
\usepackage{colortbl}
\usepackage{tcolorbox}
\usepackage{xparse}

\definecolor{iscte-iul-palette}{HTML}{14BFB8}
\usepackage{xspace}
\usepackage{amssymb}
\usepackage{enumitem}
\usepackage[percent]{overpic}
\usepackage{tablefootnote}

\usepackage{tabto}

\usepackage{academicons}

\usepackage{tikz}
\usetikzlibrary{svg.path}
\definecolor{orcidlogocol}{HTML}{A6CE39}

\tikzset{
	orcidlogo/.pic={
		\fill[orcidlogocol] svg{M 100 100 C 100 72.4 77.6 50 50 50 C 22.3 50 0 72.4 0 100 C 0 127.6 22.3 150 50 150 C 77.6 150 100 127.6 100 100 Z};
		\fill[white] svg{M 33.666 77.27 L 27.651 77.27 L 27.651 119.103 L 33.666 119.103 L 33.666 100.198 L 33.666 77.27 Z};
		\fill[white] svg{M 42.493 119.103 L 58.742 119.103 C 74.21 119.103 81.007 108.049 81.007 98.167 C 81.007 87.425 72.609 77.23 58.821 77.23 L 42.493 77.23 L 42.493 119.103 Z M 48.509 82.66 L 58.078 82.66 C 71.71 82.66 74.835 93.011 74.835 98.167 C 74.835 106.565 69.484 113.673 57.766 113.673 L 48.509 113.673 L 48.509 82.66 Z};
		\fill[white] svg{M 34.603 127.813 C 34.603 125.665 32.846 123.868 30.658 123.868 C 28.471 123.868 26.713 125.665 26.713 127.813 C 26.713 130.001 28.471 131.758 30.658 131.758 C 32.846 131.758 34.603 129.961 34.603 127.813 Z};
	}
}

\def\orcid#1{%
	\href{https://orcid.org/#1}{%
		\begin{tikzpicture}[baseline]
			\pic[scale=0.1,yshift=-7em] {orcidlogo};
		\end{tikzpicture}
	}%
}

\newcommand{\piezero}[1]{%
  \begin{tikzpicture}
    \draw (0,0) circle (1ex); \fill[white] (1ex,0) arc (0:#1:1ex) -- (0,0) -- cycle;
  \end{tikzpicture}%
}

\newcommand{\pie}[1]{%
  \begin{tikzpicture}
    \draw (0,0) circle (1ex); \fill[black!90] (1ex,0) arc (0:#1:1ex) -- (0,0) -- cycle;
  \end{tikzpicture}%
}

\newcommand{\pietwo}[1]{%
\begin{tikzpicture}[rotate=270]
 \draw (0,0) circle (1ex); \fill[black!90] (1ex,0) arc (0:#1:1ex) -- (0,0) -- cycle;
\end{tikzpicture}%
}

\newcommand{\piethree}[1]{%
\begin{tikzpicture}[rotate=180]
 \draw (0,0) circle (1ex); \fill[black!90] (1ex,0) arc (0:#1:1ex) -- (0,0) -- cycle;
\end{tikzpicture}%
}

\newcommand{\zeropct}{\piezero{0}}

\newcommand{\quarterpct}{\pie{90}}

\newcommand{\halfpct}{\pietwo{180}}

\newcommand{\threequarterpct}{\piethree{270}}

\newcommand{\fullpct}{\pie{360}}

\NewDocumentCommand{\plot}{m O{} O{12} O{H} O{trim=0cm 0cm 0cm 0cm} }{
\begin{figure}[#4]
\centering
\includegraphics[width=#3cm,clip,#5]{#1}
\caption{#2}
\label{#1}
\end{figure}
}

\newcommand{\classboard}[6]{

\begin{longtable}{p{2.4cm}cp{5.5cm}}
    \caption{#1}
     \label{table:SumContributions}
    \\\hline\noalign{\smallskip}
	 &  & \textbf{Description}\\
    \noalign{\smallskip}\hline\noalign{\smallskip}
\endfirsthead

\multicolumn{3}{c}
{ \begin{footnotesize} \tablename\ \thetable{}: continued from previous page \end{footnotesize} } \\

\hline\noalign{\smallskip}

   	&  & \textbf{Description}\\

\noalign{\smallskip}\hline\noalign{\smallskip}
\endhead

\hline \multicolumn{3}{r}{{Continued on next page}} \\ 
\noalign{\smallskip}\hline\noalign{\smallskip}
\endfoot

\hline
\endlastfoot

   \textbf{The benefit is:} &  & \\[0.2cm]
	 \textbf{Absent (0)} & \zeropct & #2\\
     \textbf{Weak (0.25)} & \quarterpct & #3\\
     \textbf{Moderate (0.5)} & \halfpct & #4\\
     \textbf{Strong (0.75)} & \threequarterpct & #5\\
     \textbf{Complete (1)} & \fullpct & #6\\

\noalign{\smallskip}\hline
\end{longtable}

}

\newcommand{\noteboard}[2]{
\scriptsize
\vspace{0.5cm}
\begin{tcolorbox}[width=\columnwidth,colback={white},title={\textbf{#1}},colbacktitle=gray!10,coltitle=black, boxrule=0.5pt, leftrule=3pt, sharp corners=all]  
    #2
\end{tcolorbox}
\normalsize
}

\newcommand{\orcidd}[1]{\href{https://orcid.org/#1}{\textcolor[HTML]{A6CE39}{\aiOrcid}}}

\begin{document}

\title{Software Development Analytics in Practice
}
\subtitle{A Systematic Literature Review}


\author{First Author         \and
        Second Author 
}

\author{João Caldeira 
\and Fernando Brito e Abreu 
\and Jorge Cardoso 
\and Rachel Simões 
\and Toacy Oliveira 
\and José Pereira dos Reis 
}


\institute{João Caldeira \textsuperscript{\orcid{0000-0003-0960-0179}} \at
              Iscte - Instituto Universitário de Lisboa, ISTAR-Iscte, Lisboa, Portugal\\
              \email{jcppc@iscte-iul.pt}           
             \and
            Fernando Brito e Abreu \textsuperscript{\orcid{0000-0002-9086-4122}} \at
              Iscte - Instituto Universitário de Lisboa, ISTAR-Iscte, Lisboa, Portugal\\
              \email{fba@iscte-iul.pt}
             \and
            Jorge Cardoso \textsuperscript{\orcid{0000-0001-8992-3466}} \at
              University of Coimbra, Coimbra, Portugal\\
              Huawei Munich Research Center, Munich, Germany\\
              \email{jcardoso@dei.uc.pt}
              \and
             Rachel Simões  \textsuperscript{\orcid{0000-0002-6046-8620}} \at
             Federal University of Rio de Janeiro, Rio de Janeiro, Brazil\\
            \email{rachelvital@cos.ufrj.br}
            \and
             Toacy Oliveira \textsuperscript{\orcid{0000-0001-8184-2442}} \at
              Federal University of Rio de Janeiro, Rio de Janeiro, Brazil\\
              \email{toacy@cos.ufrj.br}
             \and
              José Pereira dos Reis \textsuperscript{\orcid{0000-0002-2505-9565}} \at
              Iscte - Instituto Universitário de Lisboa, ISTAR-Iscte, Lisboa, Portugal\\
              \email{jvprs@iscte-iul.pt}
             }


\date{Received: date / Accepted: date}

\maketitle

\begin{abstract}
\textit{\\}

\noindent\textit{\textbf{Context}}:
Software Development Analytics is a research area concerned with providing insights to improve product deliveries and processes. Many types of studies, data sources and mining methods have been used for that purpose. \newline
\noindent\textit{\textbf{Objective}}: 
This systematic literature review aims at providing an aggregate view of the relevant studies on Software Development Analytics in the past decade, with an emphasis on its application in practical settings. \newline
\noindent\textit{\textbf{Method}}: 
Definition and execution of a search string upon several digital libraries, followed by a quality assessment criteria to identify the most relevant papers. On those, we extracted a set of characteristics (study type, data source, study perspective, development life-cycle activities covered, stakeholders, mining methods, and analytics scope) and classified their impact against a taxonomy. \newline
\noindent\textit{\textbf{Results}}: 
Source code repositories, exploratory case studies, and developers are the most common data sources, study types, and stakeholders, respectively. Testers also get moderate attention from researchers. Product managers' concerns are being addressed frequently and project managers are also present but with less prevalence. Mining methods are rapidly evolving, as reflected in their identified long list. Descriptive statistics are the most usual method followed by correlation analysis. Being software development an important process in every organization, it was unexpected to find that process mining was present in only one study. Most contributions to the software development life cycle were given in the quality dimension. Time management and costs control were less prevalent. The analysis of security aspects is even more reduced, however, evidences suggest it is an increasing topic of concern. Risk management contributions are also scarce. \newline
\noindent\textit{\textbf{Conclusions}}:
There is a wide improvement margin for software development analytics in practice. For instance, mining and analyzing the activities performed by software developers in their actual workbench, i.e., in their IDEs. Together with mining developers' behaviors, based on the evidences and trend, in a short term period we expect an increase in the volume of studies related with security and risks management.

 

\end{abstract}

%
%

\section{Introduction}
\label{sec:introduction}

Defining new processes and allocating the right resources, particularly for large organizations, is a challenging task for software project managers, primarily because it requires acquaintance on existing processes and tools, the understanding of different stakeholders, and the coordination of technical expertise in multiple domains \cite{Menzies2011TheMining}. Failing to properly manage these various aspects, namely when decisions are based on "gut feeling" (often dubbed "personal experience from past projects") may cause software development projects to produce hard to maintain technical artifacts, to surpass budget and schedule, and deliver defective products \cite{Mohagheghi2017,Emam2008}.

The Software Develpment Analytics (SDA)
research field aims at mitigating the aforementioned risks by providing the stakeholders' decision-making process with structured data-driven pieces of evidence, such as insights on software products and processes.

\subsection{Motivation}
The term ``software analytics" (SA) emerged naturally expressing the work of several research groups aiming to expand the traditional scope on analyzing software artifacts by means of mining software repositories \cite{Zhang2013SoftwarePracticeb}.
These groups conducted cutting-edge research and technology innovation in an interdisciplinary area that spans across big data, machine learning, systems, and software engineering. This approach led software practitioners to perform data exploration and analysis in order to obtain insightful and actionable information for completing various tasks around software systems, software users, and software development processes \cite{Zhang2013SoftwarePractice}.

Software Development Analytics (SDA), the adoption of analytics methods with the focus on the management of software development projects, was proposed in \cite{Buse2010AnalyticsDevelopment}. It differs from software cybernetics, which is a subdivision of ``cybernetics" in the domain of software engineering \cite{Yang2017ModernTrends}. It references the description of ``cybernetics" by Wiener, if software is regarded as part of the machine, and can be defined simply as communication and control in software. However, most researchers in the area believe software cybernetics is more diverse in scope. In fact, it is described as the interplay between software or software behaviour and control \cite{CAI2002841}. In its simplest form, the field of software cybernetics treated software problems and control problems in an integrated way \cite{Cai1173129}.

In turn, SDA is broader in its scope and based on a structured framework to identify adequate resources, ask meaningful questions, collect and analyze information properly, provide insights to the stakeholders and finally to identify the benefits for the software development life-cycle, either by looking at its past, present or future perspectives \cite{Buse2010AnalyticsDevelopment}. Although a few aspects of software cybernetics may seem to overlap with SDA, for instance, the software development activities, the mining method or the type of study, many others are missing, such as the stakeholders, the analitycs scope and more importantly, the potential contributions towards the relevant properties of software projects and where those can effectively support the decisions taken by managers.

Since the time analytics was proposed for the practice of developing software, a vast amount of literature was produced presenting stakeholders with new ways of improving the efficiency and effectiveness in developing software products, by providing insights on how to streamline the processes or to optimize resource allocation \cite{Abdellatif2015SoftwareReview}.

\subsection{Contributions}
A decade has elapsed since the first discussions on methodologies, techniques and tools to boost the adoption of analytics in the software development practice. However, there have been a small number of reports on the practice impact or benefits that software development analytics results have created on software development \cite{Zhang2013SoftwarePracticeb}. This systematic literature review (SLR) identifies, analyzes and aggregates the relevant primary studies in this period, following a well defined protocol, aligned with the best practices \cite{Kitchenham2009SystematicReview,Dyba2008StrengthEngineering}. Its main objectives are to:
\begin{itemize}
\item summarize the main types of empirical studies performed, target software life cycle activities, and corresponding data sources;
\item identify the mining methods and analytics that were applied;
\item evaluate the contributions of the selected primary studies;
\item define a taxonomy to classify the impact provided by each primary study on software development dimensions such as: quality/technical debt, time, costs, risks and security.
\end{itemize}

This paper is organized as follows: section \ref{sec:Background} provides background related to the research area and emphasizes the differences between this and previous systematic reviews in the domain. We outline the research methodology and systematic review planning in section \ref{sec:ResearchMethodology}, present the systematic review execution, data analysis, results discussion and threats to validity in section \ref{sec:DocumentingReview}, and the concluding comments appear in section \ref{sec:Conclusions}.

\section{Background}
\label{sec:Background}

Mining software repositories is currently a widespread method to gather insights from the software development process \cite{Poncin2011a,Mittal2014,Gomes2014MiningProcesses}. As these methods evolved, the software engineering practice took advantage of lessons learned and applied them in real live scenarios \cite{Menzies2015TheMethods}. The last decade has seen the birth of a multitude of analytics related companies, solutions and methodologies \cite{Menzies2015TheMethods,2016PerspectivesEngineering,Poncin2011a}, often powered by machine learning techniques. 
It was also a period where process mining saw boundless adoption in several business domains \cite{VanDerAalst2012, VanderAalst2016, Garcia2019ProcessStudy}. Both approaches, machine learning and process mining, are nowadays being used to reduce the costs of producing software products, to improve their quality, reduce time-to-market, and support the decision making-process. 


\subsection{Related Work}
\label{sec:RelatedWork}

Many SLRs have been published in the field of software engineering \cite{Kitchenham2009SystematicReview}. However, the ones addressing SDA concerns, from a holistic perspective, are scarce and often insufficiently detailed, since several aspects we deem relevant to advance the current state of the art are lacking or did not have exhaustive scrutiny. Notwithstanding, we briefly describe hereinafter all the systematic reviews whose scope somehow intersects the usual topics of SDA.

A SLR covering primary studies from 2000 to 2014, aiming to identify gaps in knowledge and open research areas in SA was presented in \cite{Abdellatif2015SoftwareReview}. It considered 19 primary studies out of 135 and the authors concluded that the practitioners who benefited most from SA studies were developers, testers, project managers (PM), portfolio managers, and higher management, with 47\% of the considered studies supporting only developers. Maintainability and reverse engineering, team collaboration and dashboards, incident management and defect prediction, the SA platform, and software effort estimation were among the domains mostly studied, with 47\% of them analyzing only one artifact.
Based on their analysis, since most of the research addresses only the low-level analytics of source code, the authors recommended researchers to use more datasets, to achieve higher confidence level in the results. They also suggested to target higher-level business decision making profiles, like portfolio management, marketing strategy, and sales directions.

A survey of the publicly available repositories and the classification of the most common ones is presented in \cite{Rodriguez2012OnProblems}. Authors also discussed the problems faced by researchers when applying machine learning or statistical techniques to them. The conclusions highlight the fact that some of the problems, such as outliers or noise, have been extensively studied in software engineering, whilst others need further research. They authors pointed out the need of further research work to deal with the imbalance and data shifting from the machine learning point of view and replication of primary studies.

A mapping study on the investigation of frequently applied empirical methods, targeted research purposes, used data sources, and applied data processing approaches and tools in empirical software engineering (ESE) was reported in \cite{Zhang2018EmpiricalSurvey}. The goal was to identify new trends and obtain interesting observations of ESE across different sub-fields of software engineering on 538 selected articles from January 2013 to November 2017. The authors observed that the trend of applying empirical methods in software engineering is continuously increasing and the most commonly applied methods are experiments, case studies and surveys, with open source projects being frequently used as data sources.  

A systematic mapping study aiming at identifying the quantity, topic, and empirical methods used, targeting the analysis of how software development practices are influenced by the use of a distributed social coding platform like GitHub, was presented in \cite{Cosentino2017AGitHub}. The authors assessed 80 publications from 2009 to 2016, and the results showed that most works focus on the interaction around coding-related tasks and project communities. They also identified some concerns about how reliable were those results based on the fact that, overall, papers used small data sets and poor sampling techniques, employed a scarce variety of methodologies and/or were hard to replicate. As a conclusion, they attested the high activity of research work around the field of open source collaboration, identified shortcomings and proposed actions to mitigate them.

A systematic mapping study providing an overview of the concerns addressed in the different phases of the software development life cycle (SDLC), was published in \cite{Dasanayake2014ConcernsStudy}. Results are reported from different viewpoints and conclusions highlight that there is a considerable variation in the use of terminologies and addressing concerns in different phases of the SDLC. 

Inspired by the increasing usage of data analytics in all areas of science and engineering, a systematic mapping study, aiming to investigate the usage of different types of analytics for software project management was presented in \cite{Nayebi2016AnalyticsGo}. The authors analyzed the accessibility of the data, as well as the degree of validation reported in the final 115 studies selected for appraisal. Results provided evidences that the majority of studies were focusing on predictive and prescriptive analytics, with almost half of the studies being essentially predictive. When comparing information versus insight as the direction of analytics, the authors found that information oriented analytics (descriptive and predictive) had a greater number of related studies (60\% of papers) than analytics searching for insight (diagnostic or prescriptive). As a final remark, their systematic mapping findings was compared with the results obtained by \cite{Buse2012InformationAnalytics}.


A systematic mapping study published in \cite{Anwar2017TowardsMapping} aims at providing an overview of the sub-domains, contribution types, research types, research methods and identify the role of software analytics in the field  of ``green software engineering". Findings show, that 163 papers out of the 260 initially found on digital libraries, used software analytical methods like statistical analysis and static analysis. Furthermore, only 11 out of the 50 papers kept for final data extraction, used software analytics techniques to foster green software engineering. Results revealed the need to develop new/improved automated software analytics tools for software practitioners, along with metrics explaining the correlation between energy usage and other quality attributes.



Our SLR aims to expand the existing knowledge about SDA, by adapting and extending the data perspectives, dimensions, and concerns identified and used by the above works. The target properties we deem as most important for a primary study to be considered relevant in this SLR are the following:

\begin{itemize}[noitemsep]

\item \textbf{Quality.} To assess the delivery of a good product or project outcome. 

\item \textbf{Scope.} To evaluate the meeting of requirements and objectives. 

\item \textbf{Time.} To track the project delivering on time.

\item \textbf{Cost.} To manage the delivery within estimated cost and effort. 

\item \textbf{Reusability.} The use of existing assets in some form within the software product development process.

\item \textbf{Maintainability.} To asses the degree to which an application is understood, repaired, or enhanced.

\item \textbf{Evolvability.} Used to describe a multifaceted quality attribute to evaluate a software system's ability to easily accommodate future changes.

\item \textbf{Performance.} To measure how effective a software system is with respect to the allocation of resources and correspondent time constraints.

\item \textbf{Security.} A cross-cutting appraise that takes into account mechanisms, such as access control, and robust design to prevent software attacks.

\item \textbf{Risk.} To address the possibility that one or more of the above properties are exposed to such levels of uncertainty that may lead them to produce undesired outcomes.
\end{itemize}

Based on this set, we propose a taxonomy to classify primary studies.

\section{Research Methodology}
\label{sec:ResearchMethodology}
In contrast to a non-structured review process, a SLR reduces bias and follows a precise and rigorous sequence of methodological steps to research literature \cite{Kitchenham2013,Wohlin2014GuidelinesEngineering}. A SLR relies on a well-defined and evaluated review protocols to search, extract, analyze, and document results as stages. This section describes the methodology applied for those activities.

\subsection{Planning the Review}
\label{subsec:PlanningReview}

\subsubsection{Research Questions}
This SLR is driven by the following research questions:


\noindent \textbf{RQ1}. \textit{What type of empirical studies have been conducted in SDA?}\\
\textbf{Justification.} The list of the main types of studies reported in SDA literature can provide a comprehensive view, both for practitioners and researchers, not only to identify areas of opportunity, but also to optimize established methods.\\

\noindent \textbf{RQ2}. \textit{What are the main data sources used for SDA related studies?}\\
\textbf{Justification.} Identifying those data sources is helpful, to provide soundness to the corresponding studies, to facilitate replication, and to stimulate the appearance of new datasets to address knowledge gaps in the field.\\

\noindent \textbf{RQ3}. \textit{What type of process/project perspective analysis was conducted?}\\
\textbf{Justification.} It refers to the ability to identify if the studies are being done before \textbf{(pre-mortem)} or after \textbf{(post-mortem)} a process/project is finished. While the latter is more frequent, namely due to the use of existing software repositories, a pre-mortem perspective can add additional value in the decision making process, as taking corrective actions on a timely manner is fundamental to keep projects or processes on track.\\

\noindent \textbf{RQ4}. \textit{What are the most studied SDLC activities?}\\
\textbf{Justification.} Understanding what SDLC activities are targeted the most (and those that are not), will help practitioners identify where most concerns and challenges are within the software development practice. It can also contribute to open new research streams to foster a deeper understanding of the complete SDLC.\\

\noindent \textbf{RQ5}. \textit{Who were the target stakeholders of these studies?}\\ 
\textbf{Justification.} Software projects are risky to conduct and continue to be difficult to predict \cite{Buse2010AnalyticsDevelopment}. SDA in practice, holds out the promise to provide decision-makers with data-driven evidences in order to better manage risk, improve efficiency and effectiveness on development projects. Studies should address the needs of different stakeholders. Identifying those beneficiaries is vital to understand if the right tools, methods and insights are reaching the ones that most need support on their daily activities.\\

\noindent \textbf{RQ6}. \textit{What are the main mining methods being used?}\\ 
\textbf{Justification.} Assessing the types of mining methods utilized helps to comprehend deeper the goals of past and current research, the limitations of their methods, benefits and conclusions  and, highlight opportunities for novel approaches in future research.\\

\noindent \textbf{RQ7}. \textit{Which type/form of analytics was applied?}\\ 
\textbf{Justification.} When exploring large volumes of data and many types of metrics, one may exploit different levels of analytics; \textbf{descriptive/diagnostics, predictive and prescriptive} \cite{Davenport2010AnalyticsResults}. Providing stakeholders in the development process with deep insights and potentially prescribing actions to take under certain circumstances is desirable. Predicting the future and prescribing actions are advanced forms of analytics which researchers and practitioners in the software development domain are expected to use.\\

\noindent \textbf{RQ8}. \textit{What were the relevant contributions to the SDLC ?}\\ 
\textbf{Justification.} On every single software development study, we should have clear benefits identified, either from using a new tool or by improving a process using a specific method. Failing to do so, reduces substantially the interest we may find in that literature and shortens the applicability of those methods in the field. SDA in practice is expected to contribute at least (but not limited to) to the following areas of concern in a software project: \textbf{technical debt/quality, costs, time, risk and security}.\\

\subsubsection{Search Strategy}
\label{search-strategy}
\noindent \textbf{Search Terms.} Based on the research questions, keywords were extracted and used to search the primary study sources. The search string included the main terms from the topics being researched, including synonyms, related items and alternative spelling. It is based on the same strategy used by \cite{Chen2011ALines} and is presented as follows:

\begin{quote}

\textit{("software analytics" OR "software development analytics")  AND ("process mining" OR "data mining" OR "big data" OR "data science") AND ("study" OR "empirical" OR "evidence based" OR "experimental" OR "in vivo")}

\end{quote}

\noindent \textbf{Digital Libraries Searched.}
A significant phase in a SLR is the search for relevant literature within the domain under study. To search for all the available literature pertinent to our research questions, in addition to some articles we added manually, the following digital libraries were queried:

\begin{itemize}[noitemsep]
 \item \href[pdfnewwindow=true]{https://dl.acm.org}{ACM Digital Library}
 \item \href[pdfnewwindow=true]{https://ieeexplore.ieee.org}{IEEE Xplore}
 \item \href[pdfnewwindow=true]{https://www.sciencedirect.com}{ScienceDirect}
 \item \href[pdfnewwindow=true]{https://www.scopus.com}{Scopus}
 \item \href[pdfnewwindow=true]{https://link.springer.com}{SpringerLink}
 \item \href[pdfnewwindow=true]{https://www.webofknowledge.com}{Web of Science }
 \item \href[pdfnewwindow=true]{https://onlinelibrary.wiley.com}{Wiley Online}
 \item \href[pdfnewwindow=true]{https://scholar.google.com}{Google Scholar}
\end{itemize}

\noindent \textbf{Publications Time Frame.} As mentioned earlier, the SDA research field emerged approximately a decade ago. Since then, as studies have gained a more structured and formal approach, it makes sense to only account for publications in journals, conferences papers, workshops and book chapters, starting from January, 1st of 2010 till the end of 2021.

\subsubsection{Selection Criteria}
\label{sec:SelectionCriteria}
We selected the above libraries based on the eagerness of collecting as many articles/papers as possible, not only because they are recognized as the most representative for Software Engineering research.
Google Scholar was selected to account for articles eventually not yet published, but also relevant to the software development domain.

\noindent\textbf{Search Stages Overview.} The outputs of the process followed to conduct the search is depicted in Figure \ref{chapter2-0-3.pdf} in appendix \ref{sec:AppendixA}. It compounds 4 sequential stages, which are described as:

\textit{\textbf{Stage 1 - Retrieve automatically results from the digital libraries}} - The referred libraries were searched using the specific syntax of each database. The search was configured in each repository to select only papers carried out within the prescribed period. The automatic search was later complemented by a manual search, according to the guidelines of Wohlin \cite{Wohlin2014GuidelinesEngineering}. 

\textit{\textbf{Stage 2 - Read titles and abstracts to identify potentially relevant studies}} - Identification of potentially relevant studies based on the analysis of title and abstract. Studies that are clearly irrelevant to the search and duplicates were discarded across the digital libraries. If there was any doubt about whether a study should be included or not, it was included for consideration in a later stage. 

\textit{\textbf{Stage 3 - Apply inclusion and exclusion criteria on reading the introduction, methods and conclusion}} - Selected studies in previous stages were reviewed, by reading the introduction, methodology section and conclusion. Afterwards, exclusion and inclusion criteria were applied as defined in Table \ref{table:TableExclusionCriteria}. At this stage, in case of doubt preventing a conclusion, the study was read in its entirety. 

\textit{\textbf{Stage 4 - Obtain primary studies and assess them}} - A list of primary studies was obtained and later submitted to critical examination using the 13 quality assessment criteria which is set out in Table \ref{table:TableQualityCriteria}.\\



\begin{table*}[!htbp] 
\caption{Exclusion and Inclusion Criteria applied at Stage 3.}
\label{table:TableExclusionCriteria}
\centering
\begin{tabular}{lp{13.5cm}}
	\hline\noalign{\smallskip}
	\textbf{Criterion} & \textbf{Description}\\
    \noalign{\smallskip}\hline\noalign{\smallskip}
	
	\multicolumn{2}{l}{\cellcolor{gray!10}\textbf{Exclusion Criteria(EC)}}\\[0.1cm]
    \textbf{EC1} & Studies published before 2010.\\
    \textbf{EC2} & Studies not written in English.\\
    \textbf{EC3} & Studies not related to the software development process.\\
    \textbf{EC4} & Studies not supported by data collected on any well designed experiment or did not use empirical data from a third party.\\
    \textbf{EC5} & Studies merely theoretical or based on expert opinion without locating a specific experience, such as: editorials, prefaces, summaries of articles, interviews, news, analysis/reviews, readers’ letters, summaries of tutorials, workshops, panels, round tables, keynotes and poster sessions.\\
    \textbf{EC6} & Studies aiming only at describing new development tools or works with the goal of simply assessing and/or validating new analytical methods without a clear statement to the benefits they may provide for the SDLC.\\\\
    
    \multicolumn{2}{l}{\cellcolor{gray!10}\textbf{Inclusion Criteria(IC)}}\\[0.1cm]
     \textbf{IC1} & Publications should be “journal” or “conference” or “workshop” or “book”.\\
     \textbf{IC2} & Works that put validated analytical methods into practice with the goal of understanding and/or improving the software development process.\\
     \textbf{IC3} & Articles that clearly addressed any of the analytics depth (RQ7) and provided benefits for the SDLC on any dimension identified in  RQ8.\\
     
    \noalign{\smallskip}\hline
\end{tabular}
\end{table*}

\subsubsection{Quality Assessment}
\label{sec:QualityAssessment}
The strategy to evaluate the quality of the studies is based on a checklist with thirteen criteria. The criteria were based on good practices for conducting
empirical research \cite{Kitchenham2009SystematicReview} and in the Critical Appraisal Skills Programme (CASP) used in different types of publications \cite{Dyba2008StrengthEngineering}.

The criteria developed to assess quality covered four main quality issues considered necessary when evaluating primary papers:

\begin{itemize} [noitemsep]
 \item \textbf{Reporting.}  Three criteria (QC1-QC3) assess if the rationale, goals and context have been clearly stated.
 \item \textbf{Rigor.} Five criteria (QC4-QC8) evaluate if a meticulous and convenient approach have been applied.
 \item \textbf{Credibility.} Two criteria (QC9-QC10) check if the findings are well presented and the gathered insights are plausible and/or credible.
 \item \textbf{Relevance.} The remain criteria (QC11-QC13) are related with the relevancy of the study for the SDLC, stakeholders and the research community.
\end{itemize}

\noindent \textbf{Selection of primary studies.} The quality of each publication should be assessed by the authors after the selection process in Stage 3. The checklist presented in Table \ref{table:TableQualityCriteria} was used to assess the credibility and thoroughness of the selected publications.
The steps that guided the selection of primary studies to reach the final results, are presented in Figure \ref{chapter2-0-3.pdf} in appendix \ref{sec:AppendixA}.

\begin{table*}[!h]
\caption{Quality Criteria.}
\label{table:TableQualityCriteria}
\centering
\begin{tabular}{lp{13.5cm}}
    \hline\noalign{\smallskip}
	\textbf{Criterion} & \textbf{Description}\\
    \noalign{\smallskip}\hline\noalign{\smallskip}
   
   \textbf{QC1} & Is the paper based on research (or merely a “lessons learned” report based on expert opinion)?\\
   \textbf{QC2} &  Is there a clear statement of the aims of the research?\\
   \textbf{QC3} & Is there an adequate description of the context in which the research was carried out?\\
  \textbf{QC4} & Was the research design appropriate to address the aims of the research?\\
  \textbf{QC5} & Was the recruitment strategy appropriate to the aims of the research?\\
   \textbf{QC6} & Was there a control group with which to compare treatments?\\
   \textbf{QC7} & Was the data collected in a way that addressed the research issue?\\
  \textbf{QC8} & Was the data analysis sufficiently rigorous?\\
  \textbf{QC9} & Has the relationship between researcher and participants been adequately considered?\\
   \textbf{QC10} & Are the datasets available to the public, thus allowing replication ?\\
    \textbf{QC11} & Is there a clear statement of findings?\\
    \textbf{QC12} & Is the study of value for research or practice?\\
    \textbf{QC13} & Did the study identified any clear benefits for the SDLC according to RQ8?\\
  
    \noalign{\smallskip}\hline
\end{tabular}
\end{table*}

Each of the 13 questions was marked as "\textbf{Yes}", "\textbf{Partially}" or "\textbf{No}". We considered a question answered as "\textbf{Partially}" in cases where we could derive relevant contents from the text, even if the details were not clearly reported.
These answers were scored as follows: "\textbf{Yes}"=\textbf{1}, "\textbf{Partially}"=\textbf{0.5}, and "\textbf{No}"=\textbf{0}. For each selected study, its quality score was computed by summing up the scores of the answers to all the quality criteria questions, being the minimum value admissible "\textbf{0}" and the maximum "\textbf{13}", in case all the questions were marked with a "\textbf{1}".

To provide validation and credibility in the quality assessment, and due to the ordinal scale for the quality criteria score, we computed using a random sample of the 173 articles, the intraclass correlation value between the raters. The results are presented later in Table \ref{table:FleissKappaSummary} in section \ref{sec:ApplyQualityCriteria}. Whenever agreement was not possible, the first author choice was taken into consideration.

\subsubsection{Data Extraction}
\label{sec:DataExtraction}
To gather standard information regarding the papers under analysis, we created a data collection form as represented in Table \ref{table:TableExtractionForm} in appendix \ref{sec:AppendixA}. This data collection form helped us to identify the date, venue and authors of the publications and also how each of them addressed the topics of our research questions.

\subsubsection{Data Synthesis}
The synthesis aimed at grouping findings from the studies in order to: identify the answers to the \textbf{RQs} presented earlier in section \ref{subsec:PlanningReview} and were organized in a spreadsheet form. This data extraction process was manually conducted by the main author. The spreadsheet was loaded and analyzed using the R statistical engine\footnote{\href[pdfnewwindow=true]{https://www.r-project.org}{https://www.r-project.org}, \href[pdfnewwindow=true]{https://rstudio.com}{https://rstudio.com}} and has now been disclosed\footnote{\href[pdfnewwindow=true]{http://dx.doi.org/10.17632/d3wdzgz88s.2}{doi:10.17632/d3wdzgz88s.2}}.

 Obtained results, plots and findings are presented and discussed in section \ref{subsec:ConductingReview}.

\subsection{Conducting the Review}
\label{subsec:ConductingReview}
This phase is responsible for executing the actions defined in section \ref{subsec:PlanningReview}. 

\subsubsection{Execute Search}
We started the review with an automatic search followed by a manual search and afterwards applied the inclusion/exclusion criteria. The search as detailed in section \ref{search-strategy}, was performed in mid 2019 and updated in the end of the last quarter of 2021, with the search string syntax being adapted to support the different search engines.
Initially we identified 3154 articles, and upon reading their titles and abstracts, the dataset was reduced to 681 articles. Following, we filtered them with the inclusion and exclusion criteria. 
Table \ref{table:InitialSearch} and Figure \ref{chapter2-0-3.pdf} in appendix \ref{sec:AppendixA}, present the summary results and workflow, respectively, for this research. 

\begin{table}[!htbp]
\caption{Digital Library Initial Search and Stages.}
\label{table:InitialSearch}
\centering
\begin{tabular}{p{3.1cm}lccccc}
	\hline\noalign{\smallskip}
	\textbf{Digital Library} & & & \multicolumn{2}{c}{\textbf{Stages}} &\\[0.1cm]
	\textbf{} & & \textbf{1} & \textbf{2} & \textbf{3} & \textbf{4}\\
    \noalign{\smallskip}\hline\noalign{\smallskip}
	
    
    
    
    \textbf{Libraries} &  & & & &\\
    
    \textbf{\tiny (ACM, Scopus, Web of Science, Science Direct, IEEE, Wiley Online, SpringerLink)} & \multirow{2}{*}{ \scriptsize 1144} & & & &\\
    

    \textbf{} & \textbf{$|-->$} & 3154 & 681 & 173 & 42\\[0.2cm]
    
    
    
    \textbf{\tiny Google Scholar and Manually Added} & \multirow{1}{*}{\scriptsize 2010} & & & & \\

    
     \noalign{\smallskip}\hline\noalign{\smallskip}
    \textbf{Total \tiny (Input for Stage 1)} & 3154 & & & &\\
    \noalign{\smallskip}\hline
\end{tabular}
\end{table}

\subsubsection{Apply Quality Assessment Criteria}
\label{sec:ApplyQualityCriteria}
The selection criteria was based on exclusions and inclusions. Table \ref{table:TableExclusionCriteria}, 
defined, in section \ref{sec:SelectionCriteria} those criteria used to assess remaining works in \textit{Stage 3}. In case of any doubt, the study was kept for analysis at a later stage.
\textit{Stage 3} provided as inputs for  \textit{Stage 4}, 173 articles, which were then assessed in their quality dimension.
At \textit{Stage 4}, we applied the quality criteria described in section \ref{sec:QualityAssessment}, resulting in 42 articles to further extract data and to answer the eight research questions.






\begin{table}[ht!]
\caption{Intraclass Correlation (ICC) (95\% Confidence Interval)}
\label{table:FleissKappaSummary}
\centering
\begin{tabular}{cccccc}
	\hline\noalign{\smallskip}
	\textbf{Subjects} & \textbf{Raters} & \textbf{ICC} & \textbf{Model} & \textbf{Type}\\
	\noalign{\smallskip}\hline\noalign{\smallskip}
     30 & 2 & 0.801 & OneWay & Agreement\\
   \noalign{\smallskip}\hline
\end{tabular}
\end{table}


We classified the studies quality level by plotting their descriptive statistics and analyzing the correspondent quartiles:

\begin{itemize}
\item \textbf{Min}: 6, \textbf{1st Q.}: 8.5, \textbf{Median}: 9.0, \textbf{Mean}: 9.007, \textbf{3rd Q.}: 9.5, \textbf{Max}: 12
\end{itemize}

As seen above,
the third quartile is at score \textbf{9.5}, therefore, we selected only the studies scoring above that mark. Based on the high level of quality, 42 studies were selected for final data extraction. Figure \ref{chapter2-1.pdf} shows the distribution of all studies per Year right after the quality assessment scoring task. 

\plot{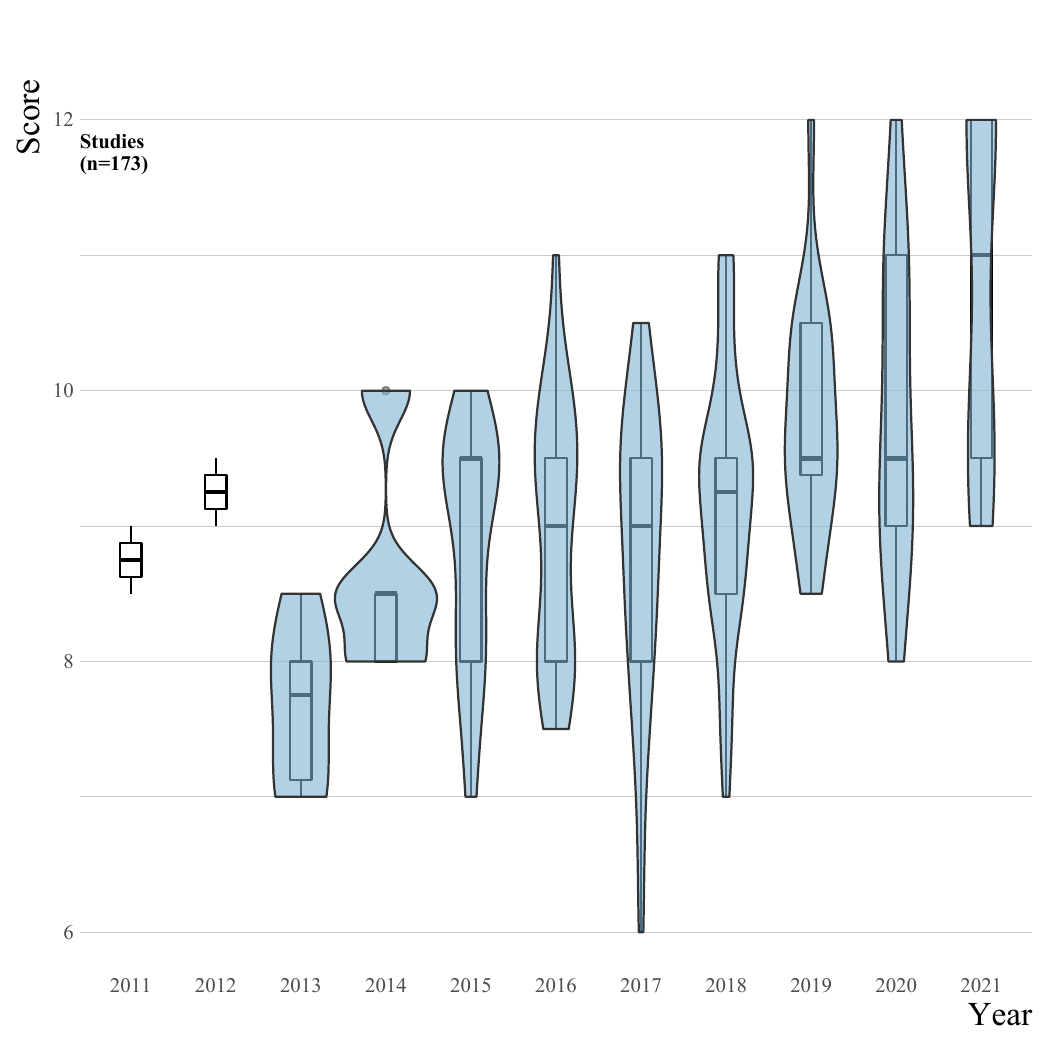}[Studies score per Year at Stage 4][8.5][H][trim=0cm 0.3cm 0cm 0.5cm]


\section{Document the Review}
\label{sec:DocumentingReview}

All selected studies and the details to support the statistics we show in section \ref{sec:Demographics}, are presented in Table \ref{tab:slr:results2} in appendix \ref{sec:AppendixB}.
In section \ref{sec:AnalysisAndFindings}, we present the main findings, comments and answers to each of the research questions. 



\subsection{Demographics}
\label{sec:Demographics}

Figure \ref{chapter2-18.pdf} shows clearly that the majority of the selected studies were published in journals. An increasing trend in publications volume is also present. 

\plot{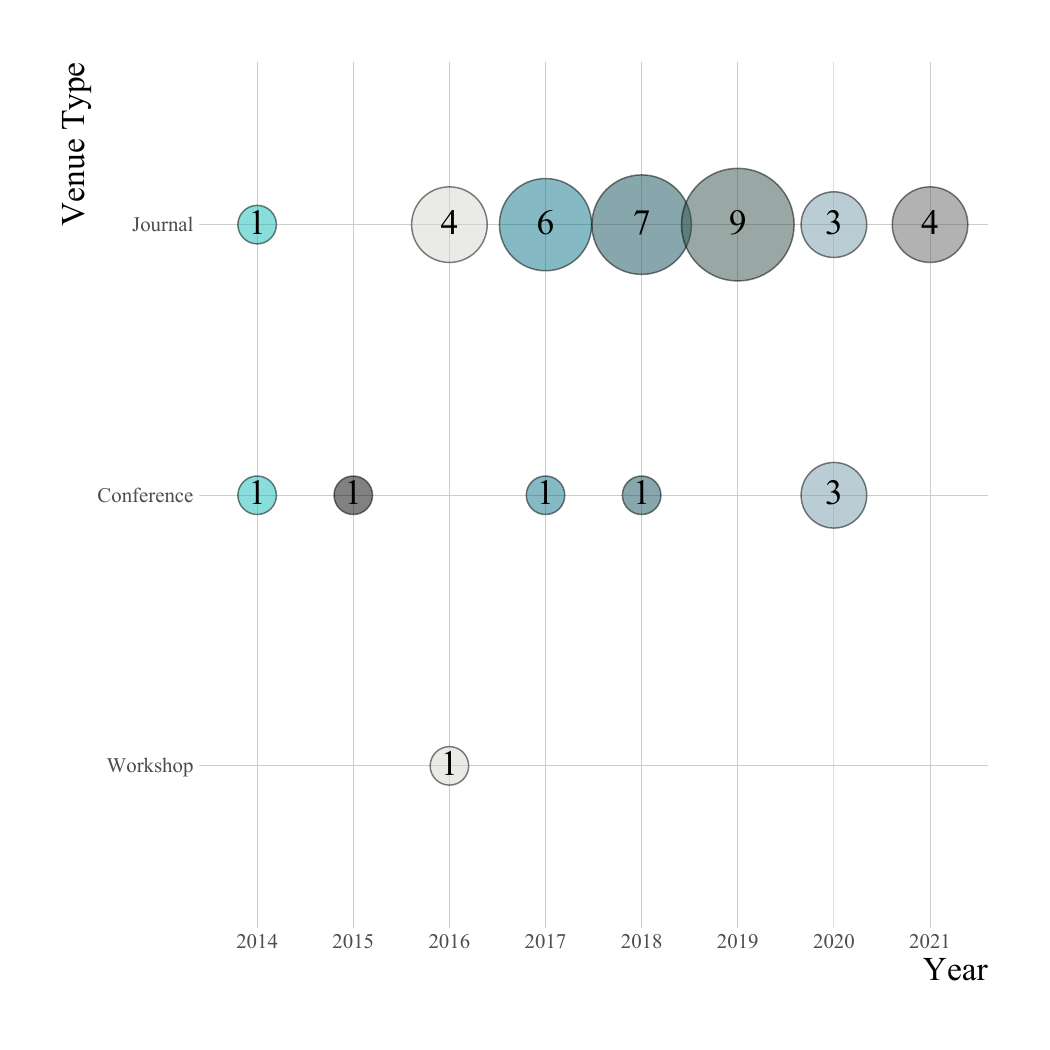}[Number of studies per Venue Type per Year][8][H][trim=1cm 1cm 1cm 1cm]

The remaining articles were published in conferences with the exception of one which comes from a workshop.
As it is possible to observe, only studies published after 2014 made the final stage of this SLR, and almost 65\% of them were published in the last 4 years. This provides some indication that, not only SDA is a relatively new practice, but also, that it is becoming mature only in the very last few years of this decade.


Looking in-depth to the publication where the studies appeared, we easily find that the Empirical Software Engineering Journal has a strong dominance among all the others.
The distribution of studies per Publication over the Years is presented in Figure \ref{chapter2-20.pdf} in appendix \ref{sec:AppendixB}. Here we can observe that only the Software Quality Journal and the Journal of Systems and Software have more than one study published within our final set of articles.


\plot{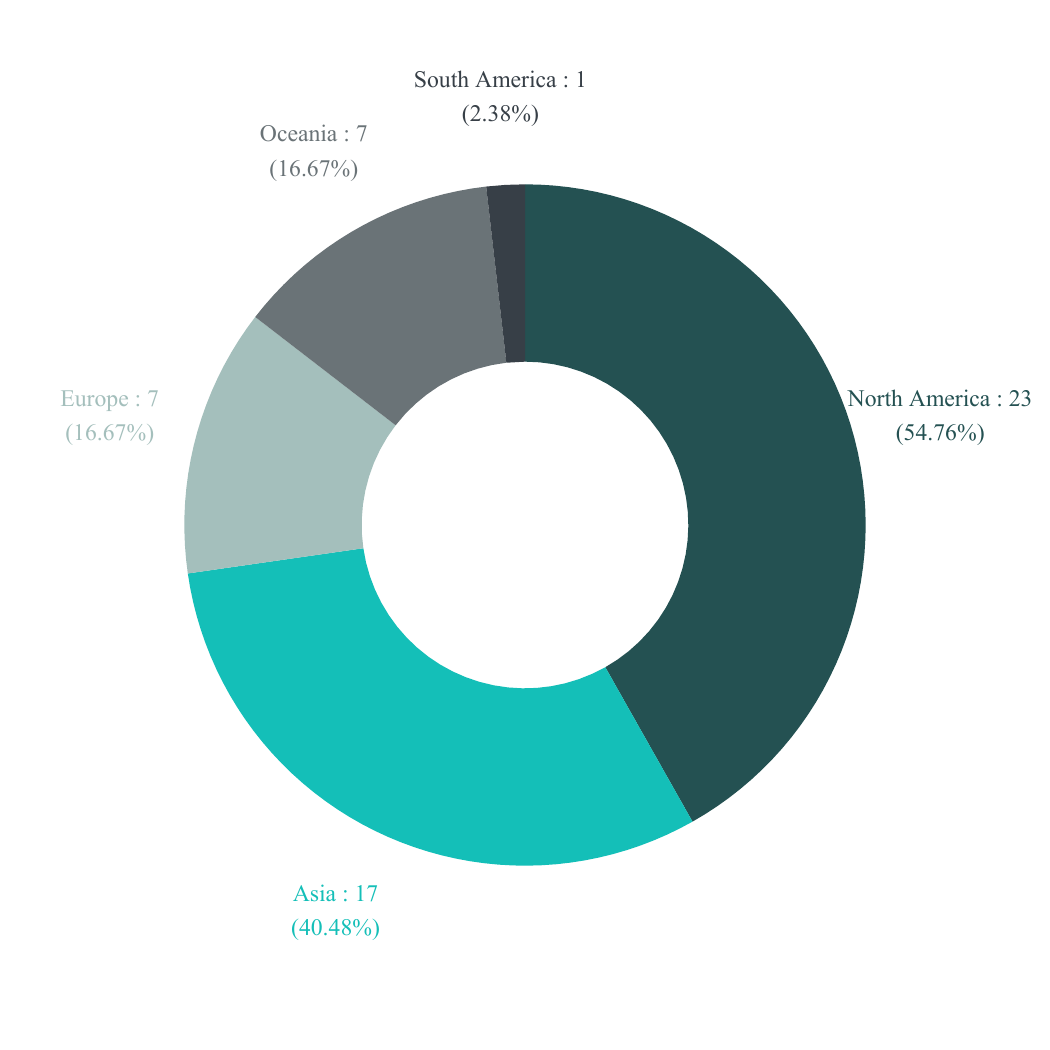}[Number of studies per Continent][8][H][trim=0.5cm 0.8cm 0.3cm 1.1cm]

North America and Asia are the most active regions researching on Software Analytics as plotted previously in Figure \ref{chapter2-35.pdf}. The collaboration between institutions from these two regions is easily detected in the large number of studies that were published in cooperation as can bee seen in Table \ref{table:SumContributors}.
Canada, Singapore, USA, Australia and China are the most effective countries in producing work in this domain. The most active institutions are also from these countries as we can observe on Figure \ref{chapter2-40.pdf}.

\plot{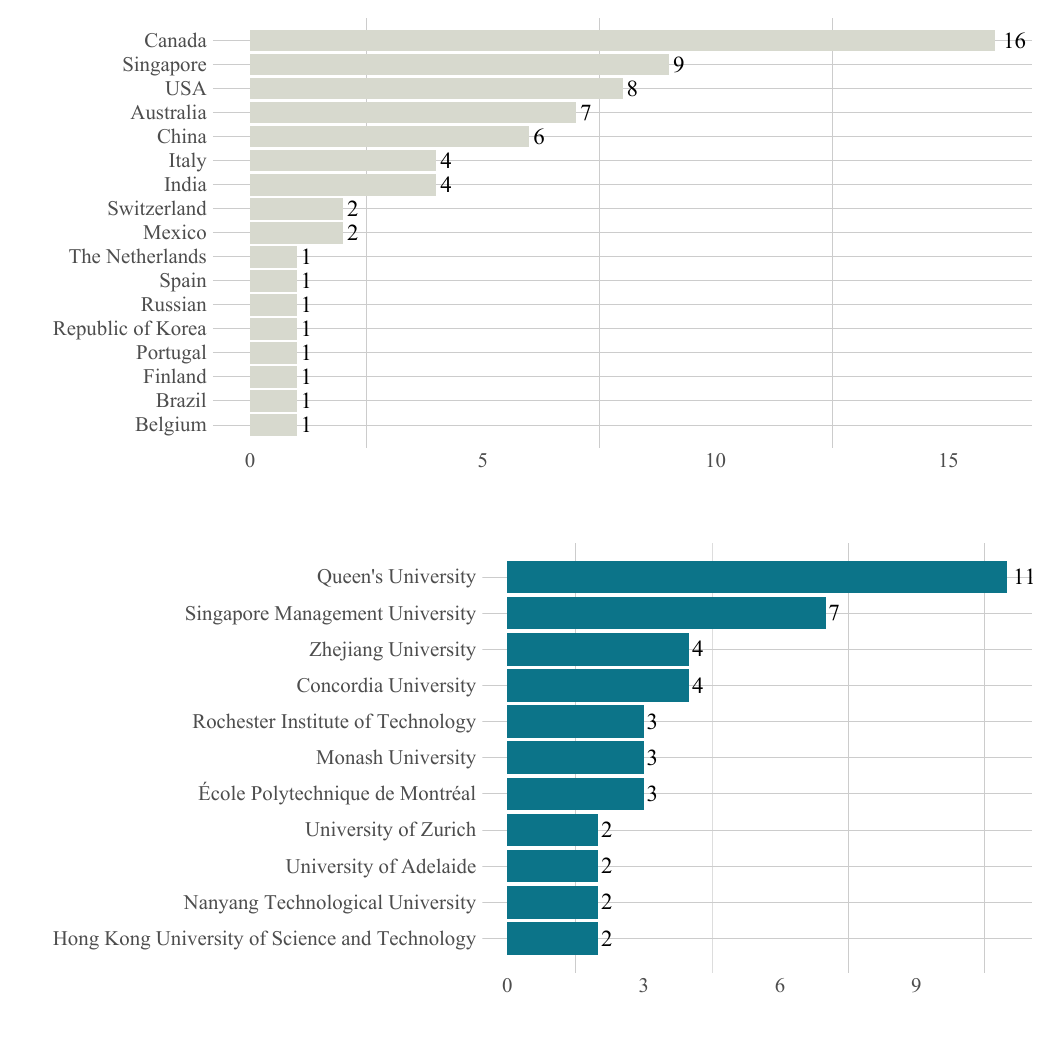}[Number of studies per Country and Institution( $>$ 1 study only)][8][H][trim=0.5cm 0.8cm 0.3cm 0cm]


Regarding authorship, which we present the details in Figure \ref{chapter2-19.pdf} in appendix \ref{sec:AppendixB}, we found that only 4 main authors appear with 2 studies in the selected papers, and one of them appear with more than one study per year. All the remaining authors are present with only one publication. This may resonate the difficulty that is to setup, document and publish such type of studies. Figure \ref{chapter2-40.pdf} in appendix \ref{sec:AppendixB}, present the frequency of contributions regarding continents, countries and institutions involved, either as primary or secondary authors, on all studies.

\subsection{Analysis and Findings}
\label{sec:AnalysisAndFindings}

It is widely accepted that we lack experimentation in Software Engineering in general. This phenomenon is even more acute on what concerns experimentation related with analytics in practice for software development. Even if this work is scarce, we should look at it collectively to try to draw some picture of the current state-of-the-art. For that purpose, a summary table with the complete information extracted to answer all the \textbf{RQs}, is presented in Table \ref{tab:slr:results} in appendix \ref{sec:AppendixC}.
In this section we present each research question and the correspondent dimension findings and their frequencies\footnote{The sum of frequencies might be bigger than the total number of selected studies(n=42) because some publications have been classified with more than one Study Type, Data Source, SDLC Activity, Stakeholder, Mining Method and/or Analytics Scope.}.
\\[0.0cm]

\subsubsection*{\textbf{RQ1. What type of empirical studies have been conducted?}}
\label{sec:RQ1}

According to the type of empirical studies provided by \cite{Zannier2006OnEngineering}, from the total number of publications, more than half, 53.12\%, are Exploratory Case Studies. Quasi-Experiments and Exploratory Case Studies combined account for 90.62\%. This is probably not a surprise, since the remaining study types are, quite often, harder to setup due to technical limitations in the data collection process or blocked by data privacy concerns raised by the entities involved. 

One publication, [S13], combines three study types: Exploratory Case Study, Quasi-Experiment and a Survey.
Having two types of empirical studies presented, we find [S31] and [S23] which combine a Exploratory Case Study and a Survey. Having a Quasi-Experiment and a Survey we have [S6] and [S24]. The remaining publications have only one empirical study type given.
Study Types found and the plot of their distribution per Year is shown on Figure \ref{chapter2-14.pdf}.

\plot{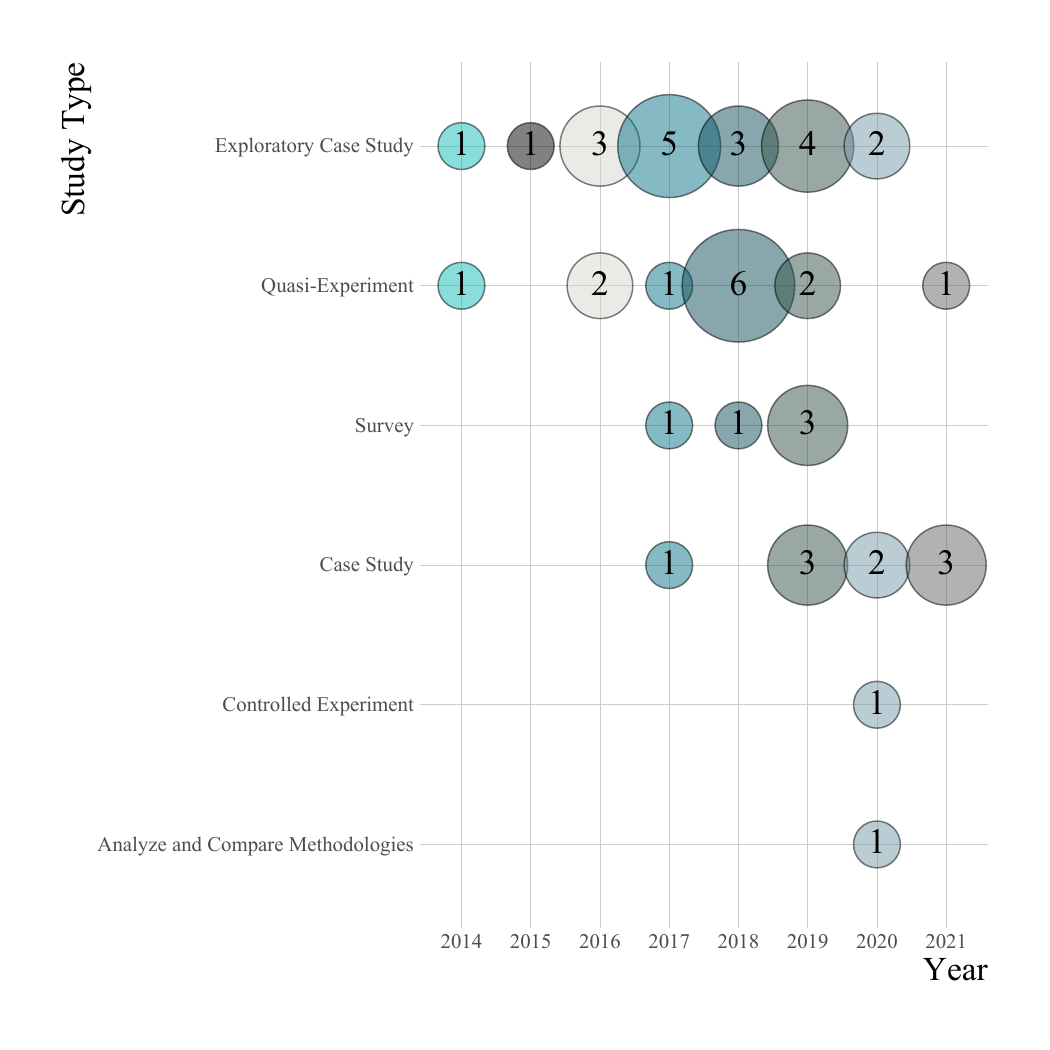}[Frequencies of Study Types per Year][8][H][trim=1cm 1cm 1cm 1cm]

No Meta-Analysis, Experience Report or Discussion had quality to reach the final stage of this SLR.
Particularly for the Controlled Experiment studies reduced presence, its worth elaborate that a controlled experiment is one in which all factors are held constant except for one: the independent variable.
It is common to compare a control group against an experimental group where all factors are identical between the two groups except for the factor being tested. This approach has the advantage that is easier to eliminate uncertainty about the significance of the results, however, it also has a considerable drawback - the effort needed to design and execute such experiments which may explain partially why there is only one study present in our final list.



\begin{table*}[!htbp]
\caption{Study Type Findings}
\label{table:SumTypes}
\centering
\begin{tabular}{lccp{9cm}}
	\hline\noalign{\smallskip}
	\textbf{Type} & \textbf{Freq.} & \textbf{Perc.} & \textbf{Ref.}\\
	\noalign{\smallskip}\hline\noalign{\smallskip}

 \textbf{Exploratory Case Study} & 19 & 45.24\% & [S04], [S05], [S09], [S10], [S13], [S16], [S20], [S22], [S23], [S25], [S26], [S27], [S28], [S29], [S30], [S31], [S32], [S37], [S40]\\[0.1cm]
\textbf{Quasi-Experiment} & 13 & 30.95\% & [S06], [S07], [S11], [S12], [S13], [S14], [S15], [S17], [S18], [S19], [S21], [S24], [S34]\\[0.1cm]
\textbf{Case Study} & 9 & 21.43\% & [S01], [S02], [S03], [S08], [S33], [S36], [S39], [S41], [S42]\\[0.1cm]
\textbf{Survey} & 5 & 11.9\% & [S06], [S13], [S23], [S24], [S31]\\[0.1cm]
\textbf{Analyze and Compare Methodologies} & 1 & 2.38\% & [S35]\\[0.1cm]
\textbf{Controlled Experiment} & 1 & 2.38\% & [S38]\\[0.1cm]
\hline

\end{tabular}
\end{table*}


We believe that sufficient conditions needed to conduct such experiments are not yet being met in software development organizations. Experiments where treatments are applied to some factors in order to later evaluate the outcomes are almost non-existent in real live scenarios. 
This may reveal that, due to revenue generation pressure, costs control and/or time restrictions,  organizations are not willing to spend time and resources to test and experiment novel approaches on analytics even when they promise potential benefits. 

\noteboard{RQ1. Highlights}{
i) Controlled Experiment studies look neglected by the community.\\
ii) 88.09\% (37/42) of works pertain to only one study type (Table \ref{tab:slr:results}).\\
iii) Evidences suggest an increasing trend in the publications quality.
}

\subsubsection*{\textbf{RQ2. What are the main data sources used for software development related studies?}}
\label{sec:RQ2}

The top four data sources: Github Repositories, Google Play Store, Git Repositories and BugZilla combined are the data sources for more than 80\% of the studies. This was somehow expected as they are generally under the public domain and contain the code, issue reports and product compilations of the most used open source projects, which are, very often used in empirical studies. This provides some evidence that the community is probably studying the most what is possible to study, simply because the datasets are under the public domain. 

Interesting to mention is the high number of publications using datasets from App Stores such as Google Play Store. This might be a relevant indicator that the researchers' focus, the profile of the end-user and the developers' characteristics are quickly and fundamentally changing.

Figure \ref{chapter2-21.pdf}, presented in appendix \ref{sec:AppendixB} plots the frequencies of all studies regarding \textbf{RQ2}.
It is proper to highlight that, from all the data sources used in more than one study, 4 are related with software configuration management systems, 2 with App Stores and each of the remaining 3 with: Bug/Issue Tracking Systems, a Q\&A Service and an Online Survey. 




\noteboard{RQ2. Highlights}{
i) Code management and bug/issue tracking systems are used frequently.\\
ii) App Stores, Q\&A services, Wikis and Forums are promising sources.\\
iii) Repositories containing developers' project interactions are scarce.
}






\subsubsection*{\textbf{RQ3. What type of process/project perspective analysis was conducted?}}
\label{sec:RQ3}

We found that all the studies were focused on a Post-Mortem approach, meaning the study was not designed to help the product/project managers take any corrective measures on a timely manner to the artifact under study. As such, any insights gathered could only impact future developments. A Post-Mortem approach provides benefits for the next product release or project, but usually, not for the one being studied as it brings no added value when proactive corrective actions are desired.

\noteboard{RQ3. Highlights}{
i) Ineffective approach to improve project under study.\\
ii) Real-time development operational support is missing.\\
iii) Worthless approach if project actions recommendation is needed.
}

\subsubsection*{\textbf{RQ4. What are the SDLC activities mostly studied?}}
\label{sec:RQ4}

According to \cite{IEEEComputerSociety2014}, in Table \ref{table:SumActivities} we summarize which activities of the SDLC, are being researched the most. Our findings show that 90.48\% and 61.9\% of the studies were targeting the Implementation and Maintenance phases, respectively. Regarding Testing, we found 13 studies. 
These results, which confirm that some phases are under-researched, require the attention of practitioners and eventually the opening of new streams of investigation on the SDLC. Software under operation was the focus of 6 studies and those were mainly related with software deployed to App Stores. Figure \ref{chapter2-13.pdf} present the statistics about all the activities studied. Table \ref{table:SumActivities} details the activities and summarizes their frequencies and identify the studies on each of them.

\begin{table*}[!htbp]
\caption{SDLC Activities Findings}
\label{table:SumActivities}
\centering
\begin{tabular}{p{3.5cm}ccp{10cm}}
	\hline\noalign{\smallskip}
	\textbf{Activity} & \textbf{Freq.} & \textbf{Perc.} & \textbf{Ref.}\\
	\noalign{\smallskip}\hline\noalign{\smallskip}

   \textbf{Implementation} & 38 & 90.48\% & [S01], [S02], [S04], [S06], [S07], [S08], [S09], [S10], [S11], [S12], [S13], [S14], [S15], [S16], [S17], [S18], [S19], [S20], [S21], [S22], [S23], [S24], [S25], [S26], [S27], [S28], [S29], [S30], [S31], [S32], [S33], [S34], [S35], [S36], [S37], [S40], [S41], [S42]\\[0.2cm]
\textbf{Maintenance} & 26 & 61.9\% & [S07], [S08], [S09], [S10], [S11], [S12], [S13], [S14], [S17], [S18], [S20], [S21], [S22], [S23], [S24], [S25], [S26], [S27], [S28], [S29], [S30], [S34], [S35], [S38], [S39], [S40]\\[0.2cm]
\textbf{Testing} & 13 & 30.95\% & [S01], [S24], [S30], [S33], [S34], [S35], [S36], [S37], [S38], [S39], [S40], [S41], [S42]\\[0.2cm]
\textbf{Debugging} & 7 & 16.67\% & [S07], [S08], [S09], [S10], [S11], [S12], [S39]\\[0.2cm]
\textbf{Operations} & 6 & 14.29\% & [S03], [S05], [S18], [S20], [S28], [S35]\\[0.2cm]
\hline

\end{tabular}
\end{table*}

\noteboard{RQ4. Highlights}{
i) More than 90\% of articles focus on the analysis of programming activities.\\
ii) Analytics for software under operation is almost non existing.\\
iii) Requirements Engineering and Design activities are not being studied.
}

\subsubsection*{\textbf{RQ5. Who were the target stakeholders of these studies?}}
\label{sec:RQ5}

All the studies targeted the Developers, and 7 were addressing Product Managers concerns.
Only 5 publications could bring any value to Testers: [S01], [S24], Educators: [S29], End-Users: [S20] and Requirements Engineers: [S18]. These findings are aligned with the results found in previous SLRs mentioned in section \ref{sec:RelatedWork}. We are predisposed to think that these results are related with the data sources also identified previously. When the majority of data sources used are product code related, it is somehow plausible that the stakeholder for that study is a developer. 
On summarizing the data about the individuals that could benefit from each study, we argue that the proper insights are not reaching all those who need support on their daily activities, namely Project Managers, Testers and Requirements Engineers. Figure \ref{chapter2-13.pdf} supports our comments by plotting the frequencies of all stakeholders targeted. 

\begin{table*}[!htbp]
\caption{Stakeholders Findings}
\label{table:SumStakeholders}
\centering
\begin{tabular}{p{3.5cm}ccp{10cm}}
	\hline\noalign{\smallskip}
	\textbf{Stakeholder} & \textbf{Freq.} & \textbf{Perc.} & \textbf{Ref.}\\
	\noalign{\smallskip}\hline\noalign{\smallskip}

 \textbf{Developers} & 42 & 100\% & [S01], [S02], [S03], [S04], [S05], [S06], [S07], [S08], [S09], [S10], [S11], [S12], [S13], [S14], [S15], [S16], [S17], [S18], [S19], [S20], [S21], [S22], [S23], [S24], [S25], [S26], [S27], [S28], [S29], [S30], [S31], [S32], [S33], [S34], [S35], [S36], [S37], [S38], [S39], [S40], [S41], [S42]\\[0.2cm]
\textbf{Product Managers} & 17 & 40.48\% & [S03], [S06], [S12], [S18], [S20], [S27], [S28], [S33], [S34], [S35], [S36], [S37], [S38], [S39], [S40], [S41], [S42]\\[0.1cm]
\textbf{Testers} & 8 & 19.05\% & [S01], [S24], [S33], [S34], [S35], [S37], [S40], [S42]\\[0.1cm]
\textbf{Project Managers} & 3 & 7.14\% & [S26], [S29], [S37]\\[0.1cm]
\textbf{Researchers} & 3 & 7.14\% & [S17], [S20], [S29]\\[0.1cm]
\textbf{Educators} & 1 & 2.38\% & [S29]\\[0.1cm]
\textbf{End-Users} & 1 & 2.38\% & [S20]\\[0.1cm]
\textbf{Requirements Engineers} & 1 & 2.38\% & [S18]\\[0.1cm]
\hline

\end{tabular}
\end{table*}

\noteboard{RQ5. Highlights}{
i) Developers keep being the main target stakeholder for SDA.\\
ii) SDA for Testers are less frequent than expected.\\
iii) High-Level management needs are not being addressed.
}

\begin{table*}[!htbp]
\caption{Mining Methods Findings}
\label{table:SumMethods}
\centering
\begin{tabular}{p{4cm}ccp{9.5cm}}
	\hline\noalign{\smallskip}
	\textbf{Stakeholder} & \textbf{Freq.} & \textbf{Perc.} & \textbf{Ref.}\\
	\noalign{\smallskip}\hline\noalign{\smallskip}

 \textbf{Descriptive Statistics} & 41 & 97.62\% & [S01], [S02], [S03], [S04], [S05], [S06], [S07], [S08], [S09], [S10], [S11], [S12], [S13], [S14], [S15], [S16], [S17], [S18], [S19], [S20], [S21], [S22], [S23], [S24], [S25], [S26], [S27], [S28], [S29], [S30], [S31], [S32], [S33], [S34], [S36], [S37], [S38], [S39], [S40], [S41], [S42]\\[0.1cm]
\textbf{Correlation Analysis} & 23 & 54.76\% & [S01], [S02], [S04], [S05], [S08], [S11], [S14], [S15], [S17], [S18], [S19], [S20], [S21], [S22], [S24], [S25], [S27], [S28], [S32], [S39], [S40], [S41], [S42]\\[0.1cm]
\textbf{Classifier Learning} & 10 & 23.81\% & [S06], [S07], [S08], [S11], [S18], [S21], [S25], [S32], [S40], [S41]\\[0.1cm]
\textbf{Pattern Extraction} & 9 & 21.43\% & [S01], [S02], [S03], [S06], [S07], [S09], [S10], [S13], [S23]\\[0.1cm]
\textbf{Hyphotesis Testing} & 8 & 19.05\% & [S04], [S05], [S11], [S14], [S15], [S17], [S18], [S39]\\[0.1cm]
\textbf{Analysis} & 4 & 9.52\% & [S33], [S34], [S35], [S38]\\[0.1cm]
\textbf{Cluster Analysis} & 4 & 9.52\% & [S20], [S22], [S25], [S32]\\[0.1cm]
\textbf{Topic Modeling} & 3 & 7.14\% & [S19], [S22], [S29]\\[0.1cm]
\textbf{Feature Extraction} & 2 & 4.76\% & [S08], [S11]\\[0.1cm]
\textbf{Redundancy Analysis} & 2 & 4.76\% & [S08], [S11]\\[0.1cm]
\textbf{Regression Models} & 2 & 4.76\% & [S19], [S20]\\[0.1cm]
\textbf{Association Rules} & 1 & 2.38\% & [S30]\\[0.1cm]
\textbf{Generalized Suffix Trees} & 1 & 2.38\% & [S32]\\[0.1cm]
\textbf{Genetic Algorithms} & 1 & 2.38\% & [S29]\\[0.1cm]
\textbf{Heuristic Features} & 1 & 2.38\% & [S07]\\[0.1cm]
\textbf{Mixed-Effect Models} & 1 & 2.38\% & [S20]\\[0.1cm]
\textbf{Natural Language Processing} & 1 & 2.38\% & [S30]\\[0.1cm]
\textbf{Process Mining} & 1 & 2.38\% & [S16]\\[0.1cm]
\hline

\end{tabular}
\end{table*}

\subsubsection*{\textbf{RQ6. What are the main mining methods being used?}}
\label{sec:RQ6}

All articles, as expected, present descriptive statistics about the domain under study. 
We know that, very often, research starts with just exploratory actions. However, understanding ``What happened" is a reduced perspective for what analytics can do for software development.
It is also not surprising that the following most frequent methods used are approaches which target the extraction of knowledge, either by correlating factors or by classifying or grouping subjects.
Hypothesis testing appears less frequently as one would expect. This may be related with the fact that all studies have, as mentioned earlier, a post-mortem approach and any results obtained are not to be used immediately to perform any corrections in the studied project. 
If used properly, that is what hypothesis testing may bring in advanced forms of analtyics.

Being software development a process, one would expect to find Process Mining methods often in the assessed studies. 
Looking deep into the data,
we can confirm that it does not hold true, which may reveal that practitioners are studying processes without the proper plethora of methods and tools. Figure \ref{chapter2-22.pdf} provide evidences for the most used mining methods.

\noteboard{RQ6. Highlights}{
i) Few studies try to make any predictions.\\
ii) Hypothesis Testing appear in only 7(21.88\%) of the studies.\\
iii) Only 1 study (3.12\%) used Process Mining methods and tools.
}


\subsubsection*{\textbf{RQ7. Which type/form of analytics was applied?}}
\label{sec:RQ7}

\begin{table*}[!htbp]
\caption{Analytics Scope Findings}
\label{table:SumAnalytics}
\centering
\begin{tabular}{p{3.5cm}ccp{10cm}}
	\hline\noalign{\smallskip}
	\textbf{Scope} & \textbf{Freq.} & \textbf{Perc.} & \textbf{Ref.}\\
	\noalign{\smallskip}\hline\noalign{\smallskip}
 
 \textbf{Descriptive} & 42 & 100\% & [S01], [S02], [S03], [S04], [S05], [S06], [S07], [S08], [S09], [S10], [S11], [S12], [S13], [S14], [S15], [S16], [S17], [S18], [S19], [S20], [S21], [S22], [S23], [S24], [S25], [S26], [S27], [S28], [S29], [S30], [S31], [S32], [S33], [S34], [S35], [S36], [S37], [S38], [S39], [S40], [S41], [S42]\\[0.1cm]
\textbf{Diagnostics} & 38 & 90.48\% & [S01], [S02], [S03], [S04], [S05], [S06], [S07], [S08], [S09], [S10], [S11], [S12], [S13], [S14], [S15], [S16], [S17], [S18], [S19], [S20], [S21], [S22], [S23], [S24], [S25], [S26], [S27], [S28], [S29], [S30], [S31], [S32], [S34], [S36], [S38], [S39], [S40], [S41]\\[0.1cm]
\textbf{Predictive} & 11 & 26.19\% & [S06], [S08], [S11], [S18], [S19], [S20], [S21], [S25], [S32], [S39], [S40]\\[0.1cm]
\textbf{Prescriptive} & 1 & 2.38\% & [S30]\\[0.1cm]
\hline

\end{tabular}
\end{table*}

Following the rationale in \textbf{RQ6}, we found all studies used Descriptive and Diagnostics Analytics together. It makes sense that understanding ``hat happened" is complemented with ``Why it happened". However, this observation is not fully aligned with the results mentioned in previous SLRs, namely in \cite{Nayebi2016AnalyticsGo}.
Although 28.12\% of the studies had some sort of prediction as a goal, that is not reflected in the prescriptive domain, where only 1 study, [S30] aims at suggesting stakeholders actions to improve or correct a development activity. Figure \ref{chapter2-13.pdf} presented in appendix \ref{sec:AppendixB} complements the analysis to this \textbf{RQ}.

\noteboard{RQ7. Highlights}{
i) Descriptive and Diagnostics Analytics seems to be found together.\\
ii) An increasing trend exists in predictive studies (Tables \ref{tab:slr:results2} \& \ref{tab:slr:results}).\\
iii) Management actions recommendation is not a common practice. 
}

\subsubsection*{\textbf{RQ8. What were the relevant contributions to the SDLC?}}
\label{sec:RQ8}

\label{sec:TechnicalDebt}
\noindent \textbf{Technical Debt.} All the studies had some sort of contribution to the quality dimension of software and no study was found to be classified with \textbf{``Absent"} under this realm. With \textbf{``Moderate"} contributions we find [S03], [S22], [S23], [S26], [S28], [S31], [S35], [S38], [S42]. Having a \textbf{``Strong"} impact we identify [S01], [S02], [S04], [S05], [S06], [S07], [S08], [S09], [S10], [S11], [S12], [S13], [S14], [S15], [S16], [S17], [S18], [S19], [S20], [S24], [S25], [S30], [S32], [S36], [S37], [S39], [S40], [S41]. Very few studies have \textbf{``Weak"} benefits identified [S21], [S27], [S29], [S33], [S34].\\

\label{sec:TimeManagement}
\noindent \textbf{Time Management.} The management of project times is analyzed in less than half of the studies since \textbf{54.76\%} of the studies provide no contribution under this dimension. We identify 11 studies, [S15], [S21], [S26], [S34], [S35], [S36], [S37], [S38], [S39], [S40], [S41] with \textbf{``Moderate"} contributions to manage the duration of product/project development. \textbf{``Weak"} benefits are present in 8 \textbf{(19.05\%)} studies [S01], [S02], [S08], [S11], [S18], [S19], [S23], [S30].\\

\label{sec:CostsControl}
\noindent \textbf{Costs Control.} A similar scenario happens with the control of costs as only 4 [S34], [S35], [S36], [S37] and 9 studies [S01], [S02], [S04], [S08], [S11], [S21], [S38], [S39], [S40] have \textbf{``Moderate"} and \textbf{``Weak"} contributions, respectively.\\

\label{sec:RisksAssessment}
\noindent \textbf{Risk Assessment.} Despite the fact that risk is cross-cut to all other dimensions identified in \textbf{RQ8}, we found only 4 studies, [S01], [S35], [S36], [S37], concerned exactly with the risk associated with the security within the software development process. The contributions given were \textbf{``Weak"} though. This means that \textbf{90.48\%} of the studies did not address at all any concerns involving risk management.\\

\label{sec:SecurityAnalysis}
\noindent \textbf{Security Analysis.} Regarding software security implementation and operations, we found very few studies where their main contributions were related to this domain. We found 5 studies, [S01], [S27], [S29], [S30] and [S36], where only the first one has a \textbf{``Strong"} classification regarding this contribution. Te remaining studies \textbf{(88.1\%)} did not mention or identified any benefits under this realm.\\

\noteboard{RQ8. Highlights}{
i) The software quality dimension consume most research resources.\\
ii) Time and Costs concerns are not being addressed sufficiently.\\
iii) Security and Risks matters need extra and aligned effort to evolve.
}

\subsection{Summary}
\label{sec:StudiesSummary}

Most of the works focus on the software quality dimension and other features are barely touched by practitioners. Improving or understanding better a project costs, risks and security aspects are contributions rare to find. Only two studies, [S1] and [S36], provide contributions across all the dimensions we assessed and they are essentially \textbf{``Moderate"} or \textbf{``Weak"} contributions.
No study was classified as \textbf{``Complete"} on any of the contribution areas identified for the SDLC.

Based on the evidences provided by this study, we observe that \textbf{80.9\%} (34 out of 42) of the studies were published in Journals, being the Empirical Software Engineering the one with more publications, 24 \textbf{(57.1\%)}.
North America and Asia are the most active regions researching on Software Analytics as plotted previously in Figure \ref{chapter2-35.pdf}. The collaboration between institutions from these two regions is easily detected in the large number of studies that were published in cooperation as can bee seen in Table \ref{table:SumContributors}.
Canada, Singapore, USA, Australia and China are the most effective countries in producing work in this domain. The most active institutions are also from these countries as we can observe on Figure \ref{chapter2-40.pdf}.

Figure \ref{chapter2-13.pdf}, which supports our answers to \textbf{RQ1, RQ4, RQ5, RQ7}, plots the frequencies of studies related with the analytics depth, study types, stakeholders and SDLC activities studied.

Figure \ref{chapter2-30.pdf} renders the evaluation off all studies across the five dimensions used to answer \textbf{RQ8}. As it is clear from the plots, Technical Debt and Time are the dimensions mostly studied.
A list of all studies with a short summary, their context, methods and results are presented in \ref{sec:AppendixB}.
A holistic perspective of all the \textbf{RQs} findings is presented in \ref{sec:AppendixC}.

\onecolumn

\begin{landscape}



\plot{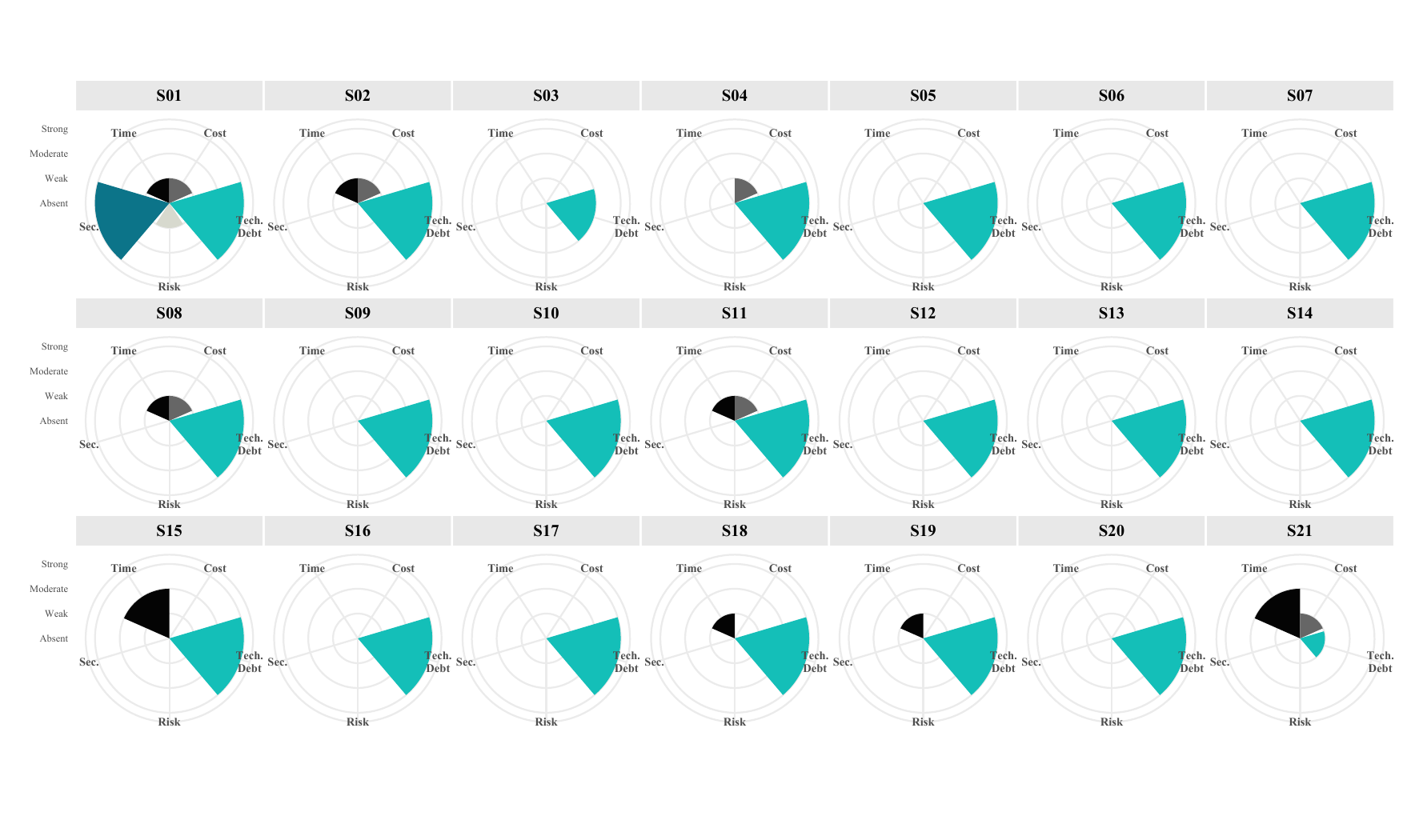}[Classification combining all 5 contribution dimensions to SDLC(RQ8)][23][H][trim=0.6cm 0.9cm 0cm 0.3cm]

\plot{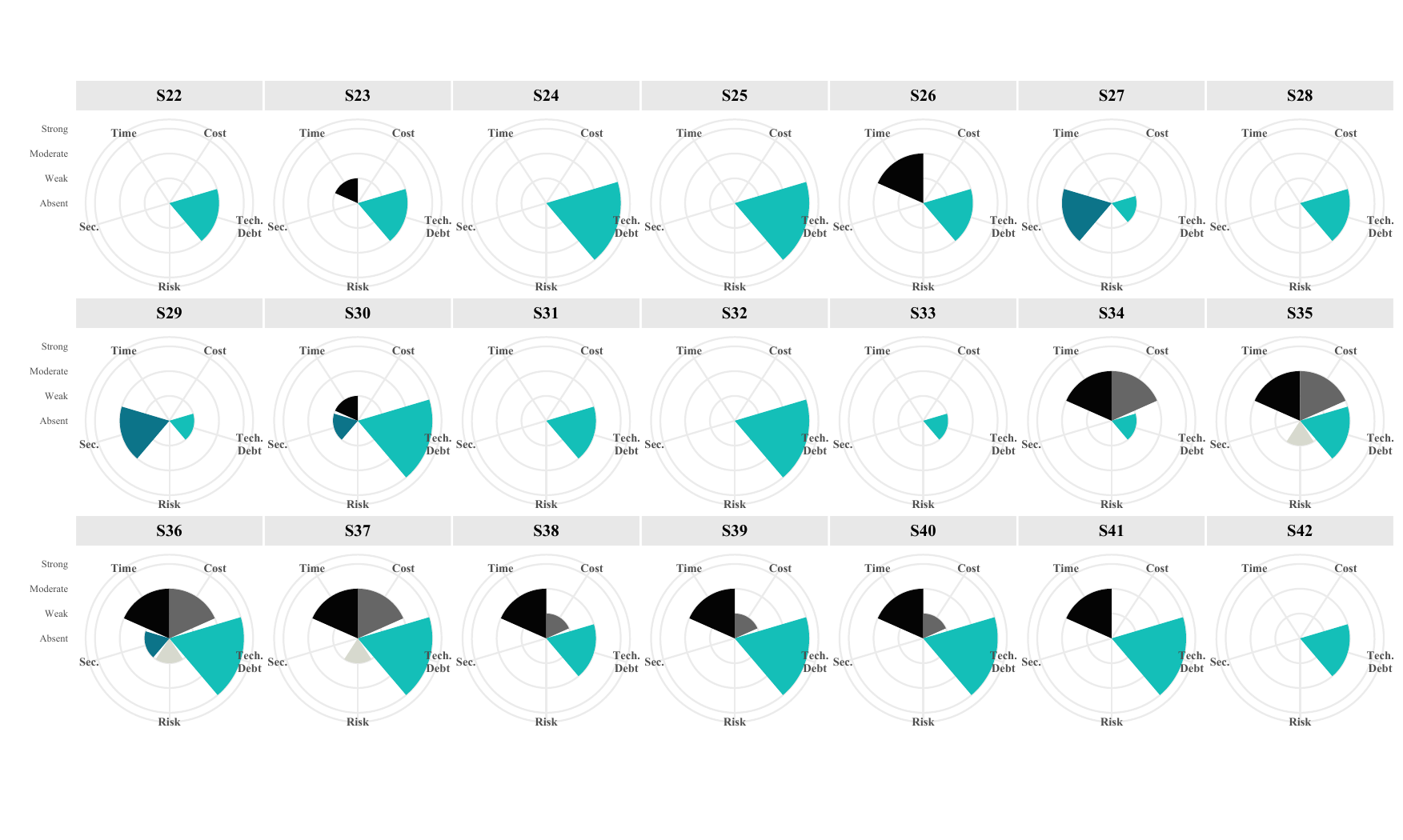}[Classification combining all 5 contribution dimensions to SDLC(RQ8)][23][H][trim=0.6cm 0.9cm 0cm 0.3cm]

\end{landscape}

\twocolumn

\subsection{Threats to validity}
\label{sec:Threatsvalidity}

The following types of validity issues were considered when interpreting the results from this review. 


\subsubsection{Construct Validity}
\label{sec:ConstructValidity}
The studies identified from the systematic review were accumulated from multiple literature databases covering relevant journals, proceedings and books. One possible threat is bias in the selection of publications. This is addressed through specifying a research protocol that defines the objectives of the study, the research questions, the search strategy and search strings used. Inclusion, exclusion criteria and blueprint for data extraction and quality assessment complements the approach to mitigate such bias.

Although supported by important literature under the software engineering domain, we followed a self-defined classification criteria for some \textbf{RQs}, specifically for \textbf{RQ8}. This method is somehow subjective as someone else might have chosen any other classification categories. 


Our dataset contains studies published until mid July, 2019. There are some evidences pointing to an increasing trend in the publishing of studies in the SDA domain, however, articles published in the second-half of 2019 which might also had good quality, were not included in this review. We excluded works where their goal was only to propose new algorithms and/or methods to analyze software development. Some of these studies had also validation experiments, however, their conclusions were related with the quality of the methods and not with any benefits potentially provided by them for the software development process. Some of those studies had also interesting approaches to improve analytics as a practice, however, they are not present in this review.

\subsubsection{Internal Validity}
\label{sec:InternalValidity}
One possible threat is the selection bias and we addressed it during the selection step of the review, i.e. the studies included in this review were identified through a thorough selection process which comprises of multiple stages. We were aiming to find high quality studies, therefore, a quality assessment was introduced and a final selection for studies ranking above the third quartile was conducted. This approach may have excluded studies with very important contributions on any of the dimensions we assessed in \textbf{RQ8} or other dimensions not covered by this review.
We used an ordinal/categorical taxonomy to assess the studies regarding \textbf{RQ8}. This classification method is still subjective and depends on the authors' contents interpretation.

\subsubsection{External Validity}
\label{sec:ExternalValidity}
There may exist other valid studies on other digital libraries which we did not search. However, we tried to reduce this limitation by exploiting the most relevant software engineering literature repositories. Studies written not in English were excluded which can also have excluded important work which otherwise would have been also mentioned.

\subsubsection{Conclusion Validity}
\label{sec:ConclusionsValidity}
There may be bias in the data extraction phase, however, this was addressed through defining a data extraction form to ensure consistent extraction of proper data to answer the research questions. We should also refer that, the findings and further comments are based on this extracted data. 
Despite the fact that high levels of validation were applied in the statistics computation of this study, there is always a small chance that any figures might be inaccurate. For this reason, we publish our final dataset to enable replication and thus allowing for further validation.

\section{Conclusions}
\label{sec:Conclusions}

We conducted a Systematic Literature Review on SDA in practice, covering a time span between 2010 and 2021. From an initial population of 3,154 papers, we kept 42 of them for appraisal. 

It targeted eight specific aspects related with the goals, sources, methods used and contributions provided in certain areas of the SDLC. Our goal was to extract the most relevant dimensions associated with software development practices and highlight where and what were the potential contributions given by those works to the SDLC. From a quality assessment perspective, our aim was also to classify the benefits provided by those studies to significant software development concerns such as: quality/technical debt, time, costs, risks and security, therefore, a taxonomy was created to evaluate them.

Source code repositories, such as GitHub and Git, and App stores like Google Play Store \textbf{(top 3 $>\textbf{50\%}$ )}, exploratory case studies ($\textbf{45.24\%}$), and developers ($\textbf{100\%}$) are the most common data sources, study types, and stakeholders, respectively. Testers ($\textbf{19.05\%}$) also get moderate attention from researchers. Product managers' ($\textbf{40.48\%}$) concerns are being addressed frequently and project managers ($\textbf{7.14\%}$) are also present but with less prevalence. Mining methods are rapidly evolving, as reflected in their identified long list. Descriptive statistics ($\textbf{97.62\%}$) are the most usual method followed by correlation analysis ($\textbf{54.76\%}$). Being software development an important process in every organization, it was unexpected to find that process mining was present in only one study ($\textbf{2.38\%}$). Most contributions to the software development life cycle were given in the quality dimension ($\textbf{100\%}$). Time management ($\textbf{45.2\%}$) and costs control ($\textbf{30.9\%}$) were less prevalent. The analysis of security aspects appear in ($\textbf{11.9\%}$) of the studies. Although with a small presence in this analysis, evidences suggest it is an increasing topic of concern. Risk management contributions are scarce ($\textbf{9.52\%}$). 


Our analysis highlighted a number of limitations and shortcomings on the SDA practice and bring the focus to open issues that need to be addressed by future research. 
It is our understanding, that our work may provide a baseline for conducting future research and the findings presented here will lead to higher quality research in this domain.

\subsection{Call for Action}
\label{sec:ConclusionsActions}
As a final remark and to trigger a call for action in the research community, the following issues should be addressed:

\begin{itemize}

\item \textbf{Repository Diversity.} 
We suggest researchers to explore different and non trivial software development related repositories, such as the IDE or other archives containing development events(eg: decisions, fine grain actions executed, etc). More and distinct datasets are expected to expand the analytics coverage on software development.

\item \textbf{Keep working on the needs of different stakeholders.}
We have evidences that the practitioners who benefit most
from the current SDA studies are the developers and many other profiles are left behind. We suggest to increase the focus on the real needs of requirements engineers, project, product and portfolio managers and higher level executives.

\item \textbf{Aim at Software Development Operational Support.} 
No studies were found providing clear evidences that the outcome of that study could benefit on a timely manner the ongoing project or product versions. 
If organizations want to focus effectively on detecting, predicting and recommending corrective actions on a timely manner, meaning, any insights gathered will have impact on current project and not solely on the next project or product version, researchers and practitioners should focus on designing advanced tools and methods to address software development operational support. 

\item \textbf{Software Development Process Mining.} Despite the fact that Process Mining is now a mature topic, almost no software process related studies uses it. We suggest its techniques and tools, to study deeper the interaction of software development stakeholders and to complement the effectiveness of assessing certain software development tasks, such as, project effort prediction, code maintenance activities and/or bug detection methods.

\item \textbf{Project Time and Costs.} We suggest more and deeper studies covering the Time and Costs of software projects. These are dimensions barely addressed by the studies we evaluated. The aforementioned topics are extremely relevant to forecast resource allocation for future projects.

\item \textbf{Address Security and Risks holistically.} Due to the unceasing digital transformation present nowadays in the society, the security of information systems will be even more critical to any organization. We now have robust methods to assess security vulnerabilities in software code. However, very little is known about the developers behaviour during the Implementation and Maintenance phases, just to name a few. Even if, in the last years, security in general became quickly a pertinent topic, the security around development processes and the involved resources are still not clearly addressed. This is a topic with increasing relevance and deserves the rapid and focused attention from the practitioners.

\item \textbf{Blockchain.} One of the most interesting, promising and relevant technological contributions to the society, was created roughly ten years ago - the birth of \textit{bitcoin} \cite{Nakamoto2009Bitcoin:System}.
Although  \textit{bitcoin} is an implementation of electronic money, it is supported by something very powerful, which can be used for many other use cases, called - \textit{blockchain} \cite{Tapscott:2016:BRT:3051781}.
The \textit{blockchain} is a mechanism which is able to keep a book of data records immutable and distributed across a multi-node network of servers. It is virtually indestructible since it has no central authority controlling it and preserves data integrity by potentially not allowing rollback on any past transactions. Additionally, if required, it guarantees that only the data owners are able to view or change their personal records and yet permit third-parties to be granted view only privileges to a selected dataset. This technology may be used embedded in SDA to anonymize and grant privacy to organizations sharing data without spoil the context associated with the development process under study.

\end{itemize}

\begin{acknowledgements}
This work was partially funded by the Portuguese Foundation for Science and Technology, under ISTAR's projects\\ UIDB/04466/2020 and UIDP/04466/2020.
\end{acknowledgements}



\bibliographystyle{spbasic}      

\bibliography{references}

\appendix

\onecolumn

\begin{landscape}

\section*{\large{Appendices}}

\section{Data Extraction}
\label{sec:AppendixA}

\section*{Selection Process}
\label{sec:selection_process}
\plot{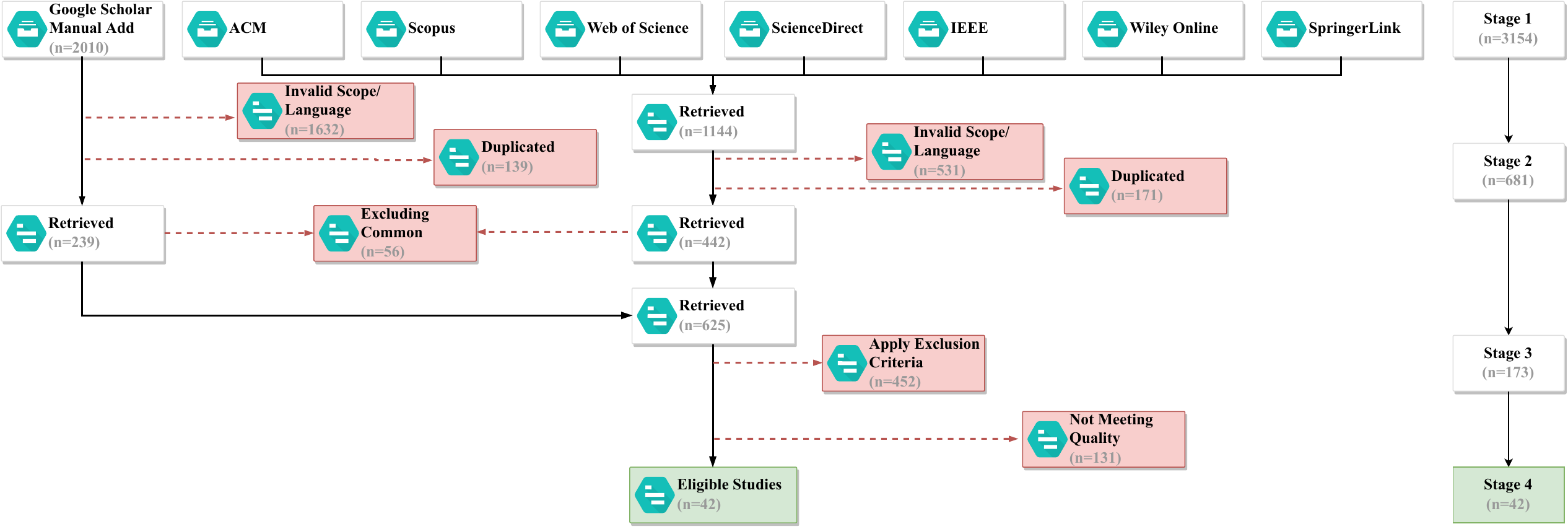}[Study Selection Process Stages][22][H][trim=0cm 0cm 0cm 0cm]

\newpage

\small

\begin{longtable}{cp{3.5cm}>{\raggedright\arraybackslash}p{14cm}>{\raggedright\arraybackslash}l}
\caption{Data Collection Form}
\label{table:TableExtractionForm}
	\\\hline\noalign{\smallskip}
	\textbf{\#} & \textbf{Item} & \textbf{Description} \\
	\noalign{\smallskip}\hline\noalign{\smallskip}
\endfirsthead

\multicolumn{3}{c}
{{ \tablename\ \thetable{}: continued from previous page}} \\
\hline\noalign{\smallskip}

	\textbf{\#} & \textbf{Item} & \textbf{Description} \\
	\noalign{\smallskip}\hline\noalign{\smallskip}
	
\endhead

\hline \multicolumn{3}{r}{{Continued on next page}} \\ 
\noalign{\smallskip}\hline\noalign{\smallskip}
\endfoot

\hline \hline
\endlastfoot

   & \multicolumn{3}{l}{\cellcolor{gray!10}\textbf{General Information}}\\[0.1cm]
   1 & Publication Id & A sequential identifier for each publication.\\
   2 & Extraction Date & Date/Time when the data was extracted.\\
   3 & Bibliography Reference & The references of each publication.\\
   4 & Publication Date & The Date/Time of publishing.\\
   5 & Publication Type & The type publication (eg: Journal, Conference, etc).\\
   6 & Publisher Name & The name of the publisher.\\
   7 & Publication Author(s) & The author(s) of the publication.\\\\
    
   & \multicolumn{3}{l}{\cellcolor{gray!10}\textbf{Addressing RQs}}\\[0.1cm]
    8 & Study Type(s) & Extracting the type of empirical study as defined in \cite{Zannier2006OnEngineering,Mohagheghi2007QualityStudies}.\\
    9 & Data Source(s) & The different types of data sources used in the publications. Admissible values are open.\\
    10 & Process Perspective & The timing of when the study was conducted (eg: \textbf{Pre-Mortem}, if study was executed before project/product was finished, \textbf{Post-Mortem}, if it was conducted after).\\
    11 & SDLC Activity(ies) & We followed the SWEBOK to build our list of admissible values \cite{IEEEComputerSociety2014}.
    \textbf{Implementation} - refers to the activity of constructing artifacts for a new product based on new defined requirements and design. \textbf{Maintenance} - refers to the task of maintaining, by changing or evolving an existing software under operation according to early defined specifications. \textbf{Testing} - refers to the automated or manual task of finding bugs and/or errors. \textbf{Debugging} - is the effort of fixing those known bugs. \textbf{Operations} - is related with the phase where the software is under exploration by the end-users. Our approach extends the taxonomy used by \cite{Dasanayake2014ConcernsStudy}.\\
    12 & Study Stakeholder(s) & The publication outcomes should be targeted to specific individuals in the software development process. We identify them here.\\
    13 & Mining Method(s) & The identification of the methods used for data mining/analysis.\\
    14 & Analytics Scope(s) & Identifies what type of analytics was performed. We used the valid options(\textbf{Descriptive Analytics}, \textbf{Diagnostics Analytics}, \textbf{Predictive Analytics} and \textbf{Prescriptive Analytics}) identified in \cite{Davenport2010AnalyticsResults}.\\
    15 & Contribution(s) to SDLC & We framed the admissible options to the following assessment dimensions of software: \textbf{Technical Debt/Quality}, \textbf{Time}, \textbf{Costs}, \textbf{Risk} and \textbf{Security}. Our approach adapt and extends some of the dimensions and concerns identified earlier in section \ref{sec:RelatedWork}.\\\\ 
    
    & \multicolumn{3}{l}{\cellcolor{gray!10}\textbf{Findings}}\\[0.1cm]
    16 & Findings and Conclusions & What were the interpretation of the results obtained.\\
    17 & Validity & Identifying the threats to the validity of the publication.\\
    18 & Relevance & What other relevant outcomes could be inferred from the publication other then the ones in item 15.\\

\end{longtable}

\end{landscape}

\begin{landscape}

\normalsize

\section{Studies List}
\label{sec:AppendixB}

\small

\begin{longtable}{p{0.3cm}ccp{3.2cm}>{\raggedright\arraybackslash}p{7cm}>{\raggedright\arraybackslash}p{5cm}}

    \caption{Systematic Literature Review Studies.}
	\label{tab:slr:results2}
    \\\hline\noalign{\smallskip}
    
	\multicolumn{1}{c}{\textbf{\shortstack[c]{\#}}} & 
	\textbf{\shortstack[c]{Score}} &
	\textbf{\shortstack[c]{Year}} &
    \textbf{\shortstack[c]{Author}} &
    \textbf{\shortstack[c]{Title}} &
	\textbf{\shortstack[c]{Publication}} \\
\noalign{\smallskip}\hline\noalign{\smallskip}
\endfirsthead

\multicolumn{6}{c}
{{ \tablename\ \thetable{}: continued from previous page}} \\
 \hline\noalign{\smallskip}

   	\multicolumn{1}{c}{\textbf{\shortstack[c]{\#}}} & 
   	\textbf{\shortstack[c]{Score}} &
	\textbf{\shortstack[c]{Year}} &
	\textbf{\shortstack[c]{Author}} &
    \textbf{\shortstack[c]{Title}} &
	\textbf{\shortstack[c]{Publication}} \\

\noalign{\smallskip}\hline\noalign{\smallskip}
\endhead

\hline \multicolumn{6}{r}{{Continued on next page}} \\ 
\noalign{\smallskip}\hline\noalign{\smallskip}
\endfoot

\hline
\endlastfoot

\textbf{S01} & 12 & 2019 & Sultana et al. & A study examining relationships between micro patterns and security vulnerabilities & Software Quality Journal \\[0.1cm]
\textbf{S02} & 10.5 & 2019 & Jha et al. & An empirical study of configuration changes and adoption in Android apps & Journal of Systems and Software \\[0.1cm]
\textbf{S03} & 10 & 2017 & Hassan et al. & An empirical study of emergency updates for top android mobile apps & Empirical Software Engineering \\[0.1cm]
\textbf{S04} & 10 & 2016 & W. Wu et al. & An exploratory study of api changes and usages based on apache and eclipse ecosystems & Empirical Software Engineering \\[0.1cm]
\textbf{S05} & 10.5 & 2017 & Taba et al. & An exploratory study on the usage of common interface elements in android applications & Journal of Systems and Software \\[0.1cm]
\textbf{S06} & 10 & 2019 & Prana et al. & Categorizing the Content of GitHub README Files & Empirical Software Engineering \\[0.1cm]
\textbf{S07} & 10 & 2018 & R. Wu et al. & ChangeLocator: locate crash-inducing changes based on crash reports & Empirical Software Engineering \\[0.1cm]
\textbf{S08} & 10.5 & 2019 & Yan et al. & Characterizing and identifying reverted commits & Empirical Software Engineering \\[0.1cm]
\textbf{S09} & 10 & 2018 & Salza et al. & Do developers update third-party libraries in mobile apps? & International Conference on Program Comprehension \\[0.1cm]
\textbf{S10} & 10 & 2019 & Liu et al. & DroidLeaks: a comprehensive database of resource leaks in Android apps & Empirical Software Engineering \\[0.1cm]
\textbf{S11} & 11 & 2018 & Fan et al. & Early prediction of merged code changes to prioritize reviewing tasks & Empirical Software Engineering \\[0.1cm]
\textbf{S12} & 10 & 2016 & McIlroy et al. & Fresh apps: an empirical study of frequently-updated mobile apps in the Google play store & Empirical Software Engineering \\[0.1cm]
\textbf{S13} & 11 & 2018 & Saborido et al. & Getting the most from map data structures in Android & Empirical Software Engineering \\[0.1cm]
\textbf{S14} & 10 & 2017 & Guerrouj et al. & Investigating the relation between lexical smells and change- and fault-proneness: an empirical study & Software Quality Journal \\[0.1cm]
\textbf{S15} & 10 & 2014 & Fucci et al. & On the role of tests in test-driven development: a differentiated and partial replication & Empirical Software Engineering \\[0.1cm]
\textbf{S16} & 10 & 2014 & Mittal et al. & Process mining software repositories from student projects in an undergraduate software engineering course & International Conference on Software Engineering \\[0.1cm]
\textbf{S17} & 10.5 & 2018 & Rakha et al. & Revisiting the performance of automated approaches for the retrieval of duplicate reports in issue tracking systems that perform just-in-time duplicate retrieval & Empirical Software Engineering \\[0.1cm]
\textbf{S18} & 10 & 2018 & Morales-Ramirez et al. & Speech-acts based analysis for requirements discovery from online discussions & Information Systems Journal \\[0.1cm]
\textbf{S19} & 10.5 & 2018 & Li et al. & Studying software logging using topic models & Empirical Software Engineering \\[0.1cm]
\textbf{S20} & 11 & 2018 & Hassan et al. & Studying the dialogue between users and developers of free apps in the Google Play Store & Empirical Software Engineering \\[0.1cm]
\textbf{S21} & 10 & 2016 & Rakha et al. & Studying the needed effort for identifying duplicate issues & Empirical Software Engineering \\[0.1cm]
\textbf{S22} & 10.5 & 2017 & Ye et al. & The structure and dynamics of knowledge network in domain-specific Q\&A sites: a case study of stack overflow & Empirical Software Engineering \\[0.1cm]
\textbf{S23} & 10.5 & 2019 & Sawant et al. & To react, or not to react: Patterns of reaction to API deprecation & Empirical Software Engineering \\[0.1cm]
\textbf{S24} & 10 & 2019 & Cruz et al. & To the attention of mobile software developers: guess what, test your app! & Empirical Software Engineering \\[0.1cm]
\textbf{S25} & 10 & 2017 & Li et al. & Towards just-in-time suggestions for log changes & Empirical Software Engineering \\[0.1cm]
\textbf{S26} & 10 & 2017 & Izquierdo-Cortazar et al. & Using Metrics to track code review performance & International Conference on Evaluation and Assessment in Software Engineering \\[0.1cm]
\textbf{S27} & 10 & 2016 & Munaiah et al. & Vulnerability severity scoring and bounties: Why the disconnect? & International Workshop on Software Analytics \\[0.1cm]
\textbf{S28} & 10 & 2015 & Tian et al. & What are the characteristics of high-rated apps? A case study on free Android Applications & International Conference on Software Maintenance and Evolution \\[0.1cm]
\textbf{S29} & 11 & 2016 & Yang et al. & What security questions do developers ask? a large-scale study of stack overflow posts & Journal of Computer Science and Technology \\[0.1cm]
\textbf{S30} & 10.5 & 2019 & Chen et al. & What's Spain's Paris ? Mining analogical libraries from Q\&A discussions & Empirical Software Engineering \\[0.1cm]
\textbf{S31} & 10 & 2017 & Jiang et al. & Why and how developers fork what from whom in GitHub & Empirical Software Engineering \\[0.1cm]
\textbf{S32} & 10.5 & 2019 & Thongtanunam et al. & Will this clone be short-lived? Towards a better understanding of the characteristics of short-lived clones & Empirical Software Engineering \\[0.1cm]
\textbf{S33} & 11 & 2020 & Avila et al. & A Data Driven Platform for Improving Performance Assessment of Software Defined Storage Solutions & Advances in Intelligent Systems and Computing \\[0.1cm]
\textbf{S34} & 12 & 2021 & Wani et al. & A Generic Analogy-Centered Software Cost Estimation Based on Differential Evolution Exploration Process & Computer Journal \\[0.1cm]
\textbf{S35} & 12 & 2020 & Rana et al. & A Study of Hyper-Parameter Tuning in the Field of Software Analytics & International Conference on Electronics, Communication and Aerospace Technology \\[0.1cm]
\textbf{S36} & 12 & 2021 & Vashisht et al. & An empirical study of heterogeneous cross-project defect prediction using various statistical techniques & International Journal of e-Collaboration \\[0.1cm]
\textbf{S37} & 10.5 & 2020 & Capizza et al. & Anomaly Detection in DevOps Toolchain & International Workshop on Software Engineering Aspects of Continuous Development and New Paradigms of Software Production and Deployment \\[0.1cm]
\textbf{S38} & 11 & 2020 & Avila et al. & Effects of contextual information on maintenance effort: A controlled experiment & Journal of Systems and Software \\[0.1cm]
\textbf{S39} & 12 & 2021 & Qu et al. & Evaluating network embedding techniques performances in software bug prediction & Empirical Software Engineering \\[0.1cm]
\textbf{S40} & 11 & 2020 & Krishna et al. & Learning actionable analytics from multiple software projects & Empirical Software Engineering \\[0.1cm]
\textbf{S41} & 10 & 2020 & Bangash et al. & On the time-based conclusion stability of cross-project defect prediction models & Empirical Software Engineering \\[0.1cm]
\textbf{S42} & 10 & 2021 & AIOmar et al. & Toward the automatic classification of Self-Affirmed Refactoring & Journal of Systems and Software \\[0.1cm]


\end{longtable}

\end{landscape}

\twocolumn

\normalsize

\section*{Comments on Studies}
\label{sec:Comments}

\small

\noindent[\textbf{S01}] explores the correlation between software vulnerabilities and code-level constructs called micro patterns. The authors analyzed the correlation between vulnerabilities and micro patterns from different viewpoints and explored whether they are related. The conclusion shows that certain micro patterns are frequently present in vulnerable classes and that there is a high correlation between certain patterns that coexist in a vulnerable class \cite{Sultana2019AVulnerabilities}.\\[0.1cm]

\noindent[\textbf{S02}] presents an empirical study to analyze commit histories of Android manifest files of hundreds of apps to understand their evolution through configuration changes. The results is a contribution to help developers in identifying change-proneness attributes, including the reasons behind the changes and associated patterns and understanding the usage of different attributes introduced in different versions of the Android platform. In summary, the results show that most of the apps extend core functionalities and improve user interface over time.  It detected that significant effort is wasted in changing configuration and then reverting back the change, and that very few apps adopt new attributes introduced by the platform and when they do, they are slow in adopting new attributes. Configuration changes are mostly influenced by functionalities extension, platform evolution and bug reports \cite{Jha2019AnApps}.\\[0.1cm]

\noindent[\textbf{S03}] studied updates in the Google Play Store by examining more than 44,000 updates of over 10,000 mobile apps, from where 1,000 were identified as emergency updates. After studying the characterirstics of the updates, the authors found that the emergency updates often have a long lifetime (i.e., they are rarely followed by another emergency update) and that updates preceding emergency updates often receive a higher ratio of negative reviews than the emergency updates \cite{Hassan2017AnApps}.\\[0.1cm]

\noindent[\textbf{S04}] analyzed and classified API changes and usages together in 22 framework releases from the Apache and Eclipse ecosystems and their client programs. The authors conclude that missing classes and methods happen more often in frameworks and affect client programs more often than the other API change types do, and that missing interfaces occur rarely in frameworks but affect client programs often. In summary, framework APIs are used on average in 35\% of client classes and interfaces and most of such usages could be encapsulated locally and reduced in number. Around 11\% of APIs usages could cause ripple effects in client programs when these APIs change. Some suggestions for developers and researchers were made to mitigate the impact of API evolution through language mechanisms and design strategies \cite{Wu2016AnEcosystems}.\\[0.1cm]

\noindent[\textbf{S05}] extracted commonly used UI elements, denoted as Common Element Sets (CESs), from user interfaces of applications. The highlight the characteristics of CESs that can result in a high user-perceived quality by proposing various metrics. From an empirical study on 1292 mobile applications, the authors observed that CESs of mobile applications widely occur among and across different categories, whilst certain characteristics of CESs can provide a high user-perceived quality. A recommendation is made, aiming  to improve the quality of mobile applications, consisting on the adoption of reusable UI templates that are extracted and summarized from CESs for developers \cite{Taba2017AnApplications}.\\[0.1cm]

\noindent[\textbf{S06}] performed a qualitative study involving the manual annotation of 4,226 README file sections from 393 randomly sampled GitHub repositories and design and evaluate a classifier and a set of features that can categorize these sections automatically. The findings show that information discussing the 'What' and 'How' of a repository hapens very often, while at the same time, many README files lack information regarding the purpose and status of a repository. A classifier was built to predict multiple categories and the F1 score obtained encourages its usage by software repositories owners. The approach presented is said to improve the quality of software repositories documentation and it has the potential to make it easier for the software development community to discover relevant information in GitHub README files \cite{Prana2019CategorizingFiles}.\\[0.1cm]

\noindent[\textbf{S07}] conducted an empirical study on characterizing the bug inducing changes for crashing bugs (denoted as crash-inducing changes). ChangeLocator was also proposed as a method to automatically locate crash-inducing changes for a given bucket of crash reports. The study approach is based on a learning model that uses features originated from the empirical study itself and a model was trained using the data from the historical fixed crashes. ChangeLocator was evaluated with six release versions of the Netbeans project. The analysis and results show that it can locate the crash-inducing changes for 44.7\%, 68.5\%, and 74.5\% of the bugs by examining only top 1, 5 and 10 changes in the recommended list, respectively, which is said to outperform other approaches \cite{Wu2018ChangeLocator:Reports}.\\[0.1cm]

\noindent[\textbf{S08}] explored if one can characterize and identify which commits will be reverted. The authors characterized commits using 27 commit features and build an identification model to identify commits that will be reverted. Reverted commits were identified by analyzing commit messages and comparing the changed content, and extracted 27 commit features that were divided into three dimensions: change, developer and message. An identification model (e.g., random forest) was built and evaluated on an empirical study on ten open source projects including a total of 125,241 commits. The findings show that the 'developer' is the most discriminative dimension among the three dimensions of features for the identification of reverted commits. However, using all the three dimensions of commit features leads to better performance of the created models \cite{Yan2019CharacterizingCommits}.\\[0.1cm]

\noindent[\textbf{S09}] conducted an empirical study on the evolution history of almost three hundred mobile apps, by investigating whether mobile developers actually update third-party libraries, checking which are the categories of libraries with respect to the developers' proneness to update their apps, looking for what are the common patterns followed by developers when updating a software library, and whether high- and low-rated apps present any particular update patterns. Results showed that mobile developers rarely update their apps with respect to the used libraries, and when they do, they mainly tend to update the libraries related to the Graphical User Interface, with the aim of keeping the mobile apps updated with the latest design trends. In some cases developers ignore updates because of a poor awareness of the benefits, or a too high cost/benefit ratio \cite{Salza2018DoApps}.\\[0.1cm]

\noindent[\textbf{S10}] extracted real resource leak bugs from a bug database named DROIDLEAKS. It consisted in mining  34 popular open-source Android apps, which resulted in a dataset having a total of 124,215 code revisions. After filtering and validating the data, the authors found, on 32 analyzed apps, 292 fixed resource leak bugs, which cover a diverse set of resource classes. To fully comprehend these bugs, they performed an empirical study, which revealed the characteristics of resource leaks in Android apps and common patterns of resource management mistakes made by developers \cite{Liu2019DroidLeaks:Apps}.\\[0.1cm]

\noindent[\textbf{S11}] built a merged code change prediction tool leveraging machine learning techniques, and extracted 34 features from code changes, which were grouped into 5 dimensions: code, file history, owner experience, collaboration network, and text. Experiments were executed on three open source projects (i.e., Eclipse, LibreOffice, and OpenStack), containing a total of 166,215 code changes. Across three datasets, the results show statistically significantly improvements in detecting merged code changes and in distinguishing important features on merged code changes from abandoned ones \cite{Fan2018EarlyTasks}.\\[0.1cm]

\noindent[\textbf{S12}] studied the frequency of updates of 10,713 mobile apps (the top free 400 apps at the start of 2014 in each of the 30 categories in the Google Play store). It was found that only $\sim$1\% of the studied apps are updated at a very frequent rate - more than one update per week and 14\% of the studied apps are updated on a bi-weekly basis (or more frequently). Results also show that 45\% of the frequently-updated apps do not provide the users with any information about the rationale for the new updates and updates exhibit a median growth in size of 6\%. The authors conclude that developers should not shy away from updating their apps very frequently, however the frequency should vary across store categories. It was observed that developers do not need to be too concerned about detailing the content of new updates as it appears that users are not too concerned about such information and, that users highly rank frequently-updated apps instead of being annoyed about the high update frequency \cite{McIlroy2016FreshStore}.\\[0.1cm]

\noindent[\textbf{S13}] studied the use of map data structure implementations by Android developers and how that relates with saving CPU, memory, and energy as these are major concerns of users wanting to increase battery life. The authors initially performed an observational study of 5713 Android apps in GitHub and then conducted a survey to assess developers' perspective on Java and Android map implementations. Finally, they performed an experimental study comparing HashMap, ArrayMap, and SparseArray variants map implementations in terms of CPU time, memory usage, and energy consumption. The conclusions provide guidelines for choosing among the map implementations: HashMap is preferable over ArrayMap to improve energy efficiency of apps, and SparseArray variants should be used instead of HashMap and ArrayMap when keys are primitive types \cite{Saborido2018GettingAndroid}.\\[0.1cm]

\noindent[\textbf{S14}] detected 29 smells consisting of 13 design smells and 16 lexical smells in 30 releases of three projects: ANT, ArgoUML, and Hibernate. Further, the authors analyzed to what extent classes containing lexical smells have higher (or lower) odds to change or to be subject to fault fixing than other classes containing design smells. The results obtained bring empirical evidence on the fact that lexical smells can make, in some cases, classes with design smells more fault-prone. In addition, it was empirically demonstrated that classes containing design smells only are more change- and fault-prone than classes with lexical smells only \cite{Guerrouj2017InvestigatingStudy}.\\[0.1cm]

\noindent[\textbf{S15}] examined the nature of the relationship between tests and external code quality as well as programmers' productivity in order to verify/refute the results of a previous study. With the focus on the role of tests, a differentiated and partial replication of the original study and related analysis was conducted. The replication involved 30 students, working in pairs or as individuals, in the context of a graduate course, and resulted in 16 software artifacts developed. Significant correlation was found between the number of tests and productivity. No significant correlation found between the number of tests and external code quality. For both cases we observed no statistically significant interaction caused by the subject units being individuals or pairs. Results obtained are consistent with the original study although, as the authors admit, there were changes in the timing constraints for finishing the task and the enforced development processes \cite{Fucci2014OnReplication}.\\[0.1cm]

\noindent[\textbf{S16}] presented an application of mining three software repositories: team wiki (used during requirement engineering), version control system (development and maintenance) and issue tracking system (corrective and adaptive maintenance) in the context of an undergraduate Software Engineering course. Visualizations, metrics and algorithms to provide an insight into practices and procedures followed during various phases of a software development life-cycle were proposed and these provided a multi-faceted view to the instructor serving as a feedback tool on development process and quality by students. Event logs produced by software repositories were mined and derived insights such as degree of individual contributions in a team, quality of commit messages, intensity and consistency of commit activities, bug fixing process trend and quality, component and developer entropy, process compliance and verification. Experimentation revealed that not only product but process quality varies signicantly between student teams and mining process aspects can help the instructor in giving directed and specific feedback. Authors, observed that commit patterns characterizing equal and un-equal distribution of workload between team members, patterns indicating consistent activity in contrast to spike in activity just before the deadline, varying quality of commit messages, developer and component entropy, variation in degree of process compliance and bug fixing quality \cite{Mittal2014ProcessCourse}.\\[0.1cm]

\noindent[\textbf{S17}] investigated the impact of the just-in-time duplicate retrieval on the duplicate reports that end up in the ITS of several open source projects, namelly Mozilla-Firefox, Mozilla-Core and Eclipse-Platform. The differences between duplicate reports for open source projects before and after the activation of this new feature were studied. Findings showed that duplicate issue reports after the activation of the just-in-time duplicate retrieval feature are less textually similar, have a greater identification delay and require more discussion to be retrieved as duplicate reports than duplicates before the activation of the feature \cite{Rakha2018RevisitingRetrieval}.\\[0.1cm]

\noindent[\textbf{S18}] exploited a linguistic technique based on speech-acts for the analysis of online discussions with the ultimate goal of discovering requirements-relevant information. The datasets used in the experimental evaluation, which are publicly available, were taken from a widely used open source software project (161120 textual comments), as well as from an industrial project in the home energy management domain. The approach used was able to successfully classify messages into Feature/Enhancement and Other, with significant accuracy. Evidence was found to support the rationale, that there is an association between types of speech-acts and categories of issues, and that there is correlation between some of the speechacts and issue priority, which could open other streams of research \cite{Morales-Ramirez2018Speech-actsDiscussions}.\\[0.1cm]

\noindent[\textbf{S19}] studied the relationship between the topics of a code snippet and the likelihood of a code snippet being logged (i.e., to contain a logging statement). The intuition driving this research, was that certain topics in the source code are more likely to be logged than others. To validate the assumptions a case study was conducted on six open source systems. The analysis gathered evidences that i) there exists a small number of "log-intensive" topics that are more likely to be logged than other topics; ii) each pair of the studied systems share 12\% to 62\% common topics, and the likelihood of logging such common topics has a statistically significant correlation of 0.35 to 0.62 among all the studied systems. In summary, the findings highlight the topics containing valuable information that can help guide and drive developers' logging decisions \cite{Li2018StudyingModels}.\\[0.1cm]

\noindent[\textbf{S20}] revisits a previous work in more depth by studying 4.5 million reviews with 126,686 responses for 2,328 top free-to-download apps in the Google Play Store. One of the major findings is that the assumption that reviews are static is incorrect. In particular, it is found that developers and users in some cases use this response mechanism as a rudimentary user support tool, where dialogues emerge between users and developers through updated reviews and responses. In addition, four patterns of developers were identified: 1) developers who primarily respond to only negative reviews, 2) developers who primarily respond to negative reviews or to reviews based on their contents, 3) developers who primarily respond to reviews which are posted shortly after the latest release of their app, and 4) developers who primarily respond to reviews which are posted long after the latest release of their app. To perform a qualitative analysis of developer responses to understand what drives developers to respond to a review, the authors analyzed a statistically representative random sample of 347 reviews with responses for the top ten apps with the highest number of developer responses. Seven drivers that make a developer respond to a review were identified, of which the most important ones are to thank the users for using the app and to ask the user for more details about the reported issue. In summary, there were significant evidences found, that it can be worthwhile for app owners to respond to reviews, as responding may lead to an increase in the given rating and that studying the dialogue between user and developer can provide valuable insights which may lead to improvements in the app store and the user support process \cite{Hassan2018StudyingStore}.\\[0.1cm]

\noindent[\textbf{S21}] empirically examined the effort that is needed for manually identifying duplicate reports in four open source projects, i.e., Firefox, SeaMonkey, Bugzilla and Eclipse-Platform. Results showed that: (i) More than 50\% of the duplicate reports are identified within half a day. Most of the duplicate reports are identified without any discussion and with the involvement of very few people; (ii) A classification model built using a set of factors that are extracted from duplicate issue reports classifies duplicates according to the effort that is needed to identify them with significant values for precision, recall and ROC area; and (iii) Factors that capture the developer awareness of the duplicate issues' peers (i.e., other duplicates of that issue) and textual similarity of a new report to prior reports are the most influential factors found. The results highlight the need for effort-aware evaluation of approaches that identify duplicate issue reports, since the identification of a considerable amount of duplicate reports (over 50\%) appear to be a relatively trivial task for developers. As a conclusion, the authors highlight the fact that, to better assist developers, research on identifying duplicate issue reports should put greater emphasis on assisting developers in identifying effort-consuming duplicate issues \cite{Rakha2016StudyingIssues}.\\[0.1cm]

\noindent[\textbf{S22}] analyzed URL sharing activities in Stack Overflow. The approach was to use open coding method to analyze why users share URLs in Stack Overflow, and develop a set of quantitative analysis methods to study the structural and dynamic properties of the emergent knowledge network in Stack Overflow. The findings show: i) Users share URLs for diverse categories of purposes. ii) These URL sharing behaviors create a complex knowledge network with high modularity, assortative mixing of semantic topics, and a structure skeleton consisting of highly recognized knowledge units. iii) The structure of the knowledge network with respect to indegree distribution is scale-free (i.e., stable), in spite of the ad-hoc and opportunistic nature of URL sharing activities, while the outdegree distribution of the knowledge network is not scale-free. iv) The indegree distributions of the knowledge network converge quickly, with small changes over time after the convergence to the stable distribution. The conclusions highlight the fact that the knowledge network is a natural product of URL sharing behavior that Stack Overflow supports and encourages, and proposed an explanatory model based on information value and preferential attachment theories to explain the underlying factors that drive the formation and evolution of the knowledge network in Stack Overflow \cite{Ye2017TheOverflow}.\\[0.1cm]

\noindent[\textbf{S23}] questioned if there was really a strong argument for the Java 9 language designers to change the implementation of the deprecation warnings feature after they notice no one was taking seriously those and continued using outdated features. The goal was to start by identifying the various ways in which an API consumer can react to deprecation and then to create a dataset of reaction patterns frequency consisting of data mined from 50 API consumers totalling 297,254 GitHub based projects and 1,322,612,567 type-checked method invocations. Findings show that predominantly consumers do not react to deprecation and a survey on API consumers was done to try to explain this behavior and by analyzing if the APIs deprecation policy had an impact on the consumers' decision to react. The manual inspection of usages of deprecated API artifacts lead to the discovery of six reaction patterns. Only 13\% of API consumers update their API versions and 88\% of reactions to deprecation is doing nothing. However the survey got a different result, where 69\% of respondents say they replace it with the recommended repalcement. Over 75\% of the API barelly affect consumers with deprecation and 15\% of the consumers are affected only by 2 APIs(hibernate-core and mongo-java-driver) \cite{Sawant2019ToDeprecation}.\\[0.1cm]

\noindent[\textbf{S24}] investigated working habits and challenges of mobile software developers with respect to testing. A key finding of this exhaustive study, using 1000 Android apps, demonstrates that mobile apps are still tested in a very ad hoc way, if tested at all. However, it is shown that, as in other types of software, testing increases the quality of apps (demonstrated in user ratings and number of code issues). Furthermore, there is evidence that tests are essential when it comes to engaging the community to contribute to mobile open source software. The authors discuss reasons and potential directions to address the findings. Yet another relevant finding of this study is that Continuous Integration and Continuous Deployment (CI/CD) pipelines are rare in the mobile apps world (only 26\% of the apps are developed in projects employing CI/CD) - authors argue that one of the main reasons is due to the lack of exhaustive and automatic testing \cite{Cruz2019ToApp}.\\[0.1cm]

\noindent[\textbf{S25}] tries to understand the reasons for log changes and, proposes an approach that can provide developers with log change suggestions as soon as they commit a code change, which is referred to as "just-in-time" suggestions for log changes. A set of measures is derived based on manually examining the reasons for log changes and individual experiences. Those measures were used as explanatory variables in random forest classifiers to model whether a code commit requires log changes. These classifiers can provide just-in-time suggestions for log changes and was evaluated with a case study on four open source projects: Hadoop, Directory Server, Commons HttpClient, and Qpid. Findings show that: i) the reasons for log changes can be grouped along four categories: block change, log improvement, dependence-driven change, and logging issue; ii) the random forest classifiers can effectively suggest whether a log change is needed; iii) the characteristics of code changes in a particular commit and the current snapshot of the source code are the most influential factors for determining the likelihood of a log change in a commit \cite{Li2017TowardsChanges}.\\[0.1cm]

\noindent[\textbf{S26}] designed and conducted, with the continuous feedback of the Xen Project Advisory Board, a detailed analysis focused on finding problems associated with the large increase over time in the number of messages related to code review. The increase was being perceived as a potential signal of problems with their code review process and the usage of metrics was suggested to track the performance of it. As a result, it was learned how in fact the Xen Project had some problems, but at the moment of the analysis those were already under control. It was found as well how diferent the Xen and Netdev projects were behaving with respect to code review performance, despite being so similar from many points of view. A comprehensive methodology, fully automated, to study Linux-style code review was proposed \cite{Izquierdo-Cortazar2017UsingPerformance}.\\[0.1cm]

\noindent[\textbf{S27}] analyzed the Common Vulnerability Scoring System (CVSS) scores and bounty awarded for 703 vulnerabilities across 24 products. CVSS is the de facto standard for vulnerability severity measurement today and is crucial in the analytics driving software fortification. It was found a weak correlation between CVSS scores and bounties, with CVSS being more likely to underestimate bounty. Such a negative result is suggested to be a cause for concern. The authors, investigated why the measurements were so discordant by i) analyzing the individual questions of CVSS with respect to bounties and ii) conducting a qualitative study to find the similarities and diferences between CVSS and the publicly-available criteria for awarding bounties. It was found that the bounty criteria were more explicit about code execution and privilege escalation whereas CVSS makes no explicit mention of those. Another lesson learnt was that bounty valuations are evaluated solely by project maintainers, whereas CVSS has little provenance in practice \cite{Munaiah2016VulnerabilityDisconnect}.\\[0.1cm]

\noindent[\textbf{S28}] through a case study on 1,492 high-rated and low-rated free apps mined from the Google Play store, investigated 28 factors along eight dimensions to understand how high-rated apps are different from low-rated apps. The search for the most influential factors was also addressed by applying a random-forest classifier to identify high-rated apps. The results show that high-rated apps are statistically significantly different in 17 out of the 28 factors that we considered. The experiment also presents eveidences for the fact that the size of an app, the number of promotional images that the app displays on its web store page, and the target SDK version of an app are the most influential factors \cite{Tian2015WhatApplications}.\\[0.1cm]

\noindent[\textbf{S29}] conducted a large-scale study on security-related questions on Stack Overflow. Two heuristics were used to extract from the dataset the questions that are related to security based on the tags of the posts. Later, to cluster different security-related questions based on their texts, an advanced topic model, Latent Dirichlet Allocation (LDA) tuned using Genetic Algorithm (GA) was used. Results show that security-related questions on Stack Overflow cover a wide range of topics, which belong to five main categories: web security, mobile security,  cryptography, software security, and system security. Among them, most questions are about web security. In addition, it was found that the top four most popular topics in the security area are "Password", "Hash", "Signature" and "SQL Injection", and the top eight most difficulty security-related topics are "JAVA Security", "Asymetric Encryption", "Bug", "Browser Security", "Windows Authority", "Signature", "ASP.NET" and "Password", suggesting these are the ones in need for more attention \cite{Yang2016WhatPosts}.\\[0.1cm]

\noindent[\textbf{S30}] present an approach to recommend analogical libraries based on a knowledge base of analogical libraries mined from tags of millions of Stack Overflow questions. The approach was implemented in a proof-of-concept web application and more than 34.8 thousands of users visited the website from November 2015 to August 2017. Results show evidences that accurate recommendation of analogical libraries is not only possible but also a desirable solution. Authors validated the usefulness of their analogical-library recommendations by using them to answer analogical-library questions in Stack Overflow \cite{Chen2019WhatsDiscussions}.\\[0.1cm]

\noindent[\textbf{S31}] explored why and how developers fork what from whom in GitHub. This approach was supported by collecting a dataset containing 236,344 developers and 1,841,324 forks. It was also validated by a survey in order to analyze the programming languages and owners of forked repositories. Among the main findings we have: i) Developers fork repositories to submit pull requests, fix bugs, add new features and keep copies etc. Developers find repositories to fork from various sources: search engines, external sites (e.g., Twitter, Reddit), social relationships, etc. More than 42\% of developers that were surveyed agree that an automated recommendation tool is useful to help them pick repositories to fork, while more than 44.4\% of developers do not value a recommendation tool. Developers care about repository owners when they fork repositories. ii) A repository written in a developers' preferred programming language is more likely to be forked. iii) Developers mostly fork repositories from creators. In comparison with unattractive repository owners, attractive repository owners have higher percentage of organizations, more followers and earlier registration in GitHub. The results show that forking is mainly used for making contributions of original repositories, and it is beneficial for OSS community. In summary, there is evidence of the value of recommendation and provide important insights for GitHub to recommend repositories \cite{Jiang2017WhyGitHub}.\\[0.1cm]

\noindent[\textbf{S32}] designed and executed an empirical study on six open source Java systems to better understand the life expectancy of clones. A random forest classifier was built with the aim of determining the life expectancy of a newly-introduced clone (i.e., whether a clone will be short-lived or longlived) and it was confimed to have good accuracy on that task. Results show that a large number of clones (i.e., 30\% to 87\%) lived in the systems for a short duration. Moreover, it finds that although short-lived clones were changed more frequently than long-lived clones throughout their lifetime, short-lived clones were consistently changed with their siblings less often than long-lived clones.  Findings show that the churn made to the methods containing a newly-introduced clone, the complexity and size of the methods containing the newly- introduced clone are highly influential in determining whether the newly-introduced clone will be short-lived. Furthermore, the size of a newly-introduced clone shares a positive relationship with the likelihood that the newly introduced clone will be short-lived. Results suggest that, to improve the efficiency of clone management efforts, such as the planning of the most effective use of their clone management resources in advance, practitioners can leverage the presented classifiers and insights in order to determine the life expectancy of clones \cite{Thongtanunam2019WillClones}.\\[0.1cm]

\noindent[\textbf{S33}] This paper introduces DDP (Data Driven Plataform) platform, a scalable platform to analyze and exploit performance data. This platform centralizes, analyzes and visualizes the performance data produced during the software development cycle. DDP employs big data and analytics technology to collect, store and process performance data in an efficient and integrated way. They have demonstrated the successful application of DDP for Spectrum Scale, a software defined storage solution, where they have been able to implement performance regression data analysis to validate the performance consistency of new produced builds \cite{Avila2020DataDrivenPlataform}.\\[0.1cm]

\noindent[\textbf{S34}] To help the industry practitioners in these situations, a analogy-centered model based on differential evolution exploration process is proposed in this research study. The proposed model has been assessed on 676 projects from 5 different data sets and the results achieved are significantly better when compared with other benchmark analogy-based estimation studies \cite{Wani2020DevOps}.\\[0.1cm]

\noindent[\textbf{S35}] The paper attempts to analyze and compare various methodologies to tune the defect predictors. The research papers which are analyzed here have used data-set from the PROMISE repository, open-source \cite{Rana2020SoftwareAnalytics}.\\[0.1cm]

\noindent[\textbf{S36}] This paper evaluates empirically and theoretically heterogeneous Cross-project defect prediction (HCPDP) modeling, which comprises of three main phases: Feature ranking and feature selection, metric matching, and finally, predicting defects in the target application. The research work has been experimented on 13 benchmarked datasets of three open source projects. Results show that performance of HCPDP is very much comparable to baseline within project defect prediction \cite{Vashisht2021StatisticalTechnique}.\\[0.1cm]

\noindent[\textbf{S37}] An anomaly detection system can operate in the staging environment to compare the current incoming release with previous ones according to predefined metrics. The analysis is conducted before going into production to identify anomalies. In this paper, they describe a prototypical implementation of the aforementioned idea in the form of a proof-of-concept \cite{Capizzi2020DevOps}.\\[0.1cm]

\noindent[\textbf{S38}] This article reports a controlled experiment that compares the effort to implement changes, the correctness and the maintainability of an existing application between two projects; one that uses qualitative dashboards depicting contextual information, and one that does not \cite{Avila2020Maintenance}.\\[0.1cm]

\noindent[\textbf{S39}] In this paper conducts an extensive empirical study to evaluate network embedding algorithms in bug prediction by utilizing and extending node2defect, a newly proposed bug prediction model that combines the embedded vectors with traditional software engineering metrics through concatenation. Experiments are conducted based on seven network embedding algorithms,two effort-aware models, and 13 open-source Java systems \cite{Qu2021BugPrediction}.\\[0.1cm]

\noindent[\textbf{S40}] This paper presents a technology for prescriptive software analytics. Their planner offers users a guidance on what action to take in order to improve the quality of a software project. Our preferred planning tool is BELLTREE, which performs cross-project planning with encouraging results.With our BELLTREE planner, we show that it is possible to reduce several hundred defects in software projects \cite{Krishna2020PredictingSoftwareDefect}.\\[0.1cm]

\noindent[\textbf{S41}] In this paper they investigate whether conclusions in the area of defect prediction, if the claims of the researchers are stable throughout time. This case study provides evidence that in the field of defect prediction the context of evaluation (in our case, time) plays an important role \cite{Bangash2020DefectPrediction}.\\[0.1cm]

\noindent[\textbf{S42}] In this paper, they propose a two-step approach to first identify whether a commit describes developer-related refactoring events, then to classify it according to the refactoring common quality improvement categories \cite{AlOmar2021Refactoring}.\\[0.1cm]

\newpage
\onecolumn

\section*{General Statistics}
\plot{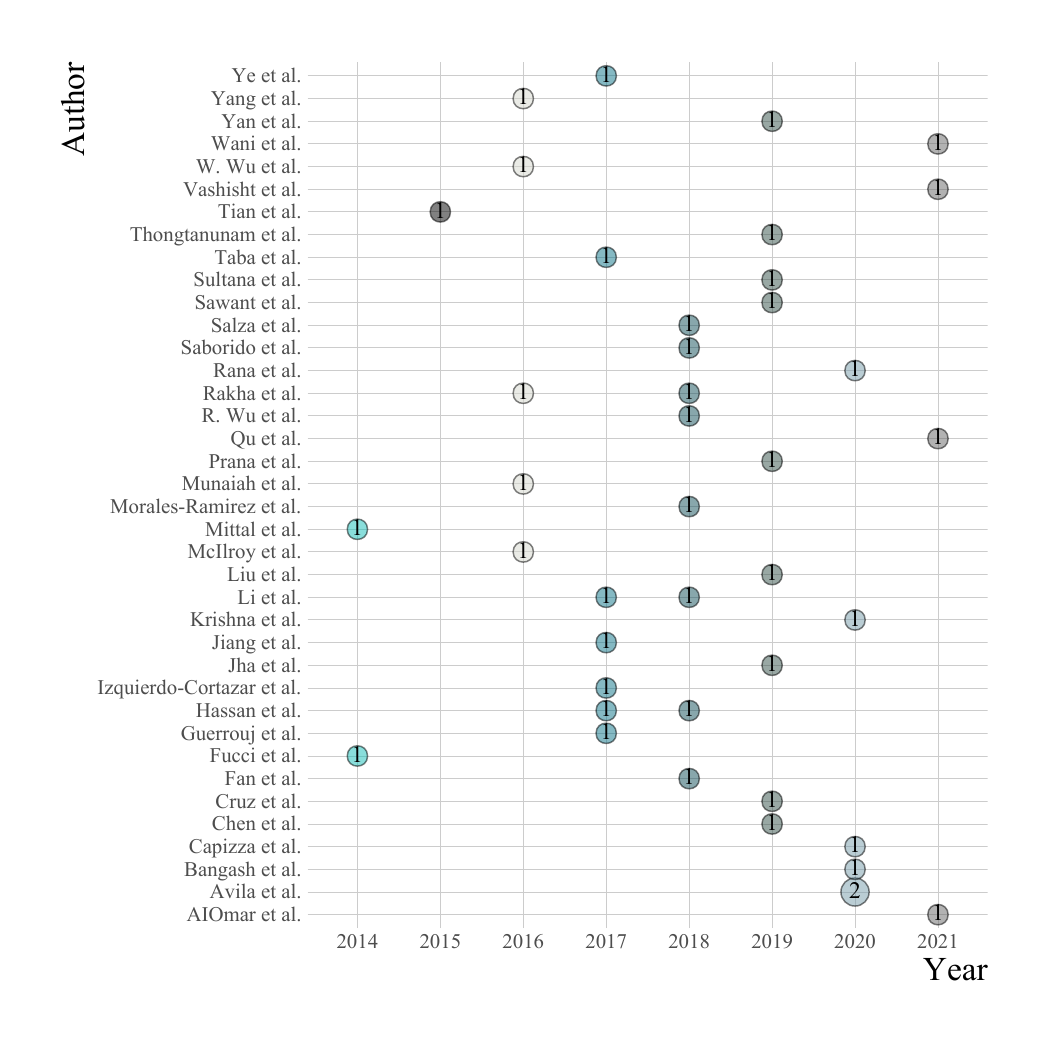}[Number of studies published by each main author over the years][15][!htbp][trim=1cm 1cm 1cm 1cm]




\footnotesize


\begin{longtable}{p{5.1cm}ccp{7.5cm}}
    \caption{List of all Contributors}
     \label{table:SumContributors}
    \\\hline\noalign{\smallskip}
		\textbf{Name} & \textbf{Freq.} & \textbf{Perc.} & \textbf{Ref.}\\
    \noalign{\smallskip}\hline\noalign{\smallskip}
\endfirsthead

\multicolumn{4}{c}
{ \begin{footnotesize} \tablename\ \thetable{}: continued from previous page \end{footnotesize} } \\

\hline\noalign{\smallskip}

   		\textbf{Name} & \textbf{Freq.} & \textbf{Perc.} & \textbf{Ref.}\\

\noalign{\smallskip}\hline\noalign{\smallskip}
\endhead

\hline \multicolumn{4}{r}{{Continued on next page}} \\ 
\noalign{\smallskip}\hline\noalign{\smallskip}
\endfoot

\hline
\endlastfoot

   \textbf{Ahmed E. Hassan} & 10 & 23.81\% & [S03], [S08], [S12], [S17], [S19], [S20], [S21], [S25], [S28], [S32]\\
\textbf{David Lo} & 7 & 16.67\% & [S06], [S08], [S11], [S24], [S28], [S29], [S31]\\
\textbf{Weiyi Shang} & 5 & 11.9\% & [S03], [S19], [S21], [S25], [S32]\\
\textbf{Xin Xia} & 4 & 9.52\% & [S08], [S11], [S29], [S31]\\
\textbf{Foutse Khomh} & 3 & 7.14\% & [S04], [S13], [S14]\\
\textbf{Giuliano Antoniol} & 3 & 7.14\% & [S04], [S13], [S14]\\
\textbf{Yann-Gael Guéhéneuc} & 3 & 7.14\% & [S04], [S13], [S14]\\
\textbf{Cor-Paul Bezemer} & 2 & 4.76\% & [S17], [S20]\\
\textbf{Heng Li} & 2 & 4.76\% & [S19], [S25]\\
\textbf{Mohamed Sami Rakha} & 2 & 4.76\% & [S17], [S21]\\
\textbf{Safwat Hassan} & 2 & 4.76\% & [S03], [S20]\\
\textbf{Shanping Li} & 2 & 4.76\% & [S08], [S11]\\
\textbf{Shing-Chi Cheung} & 2 & 4.76\% & [S07], [S10]\\
\textbf{Ying Zou} & 2 & 4.76\% & [S05], [S25]\\
\textbf{Zhenchang Xing} & 2 & 4.76\% & [S22], [S30]\\
\textbf{Abdul Ali Bangash} & 1 & 2.38\% & [S41]\\
\textbf{Abram Hindle} & 1 & 2.38\% & [S41]\\
\textbf{Ajay Kumar Jha} & 1 & 2.38\% & [S02]\\
\textbf{Alberto Bacchelli} & 1 & 2.38\% & [S23]\\
\textbf{Ali Ouni} & 1 & 2.38\% & [S42]\\
\textbf{Anand Ashok Sawant} & 1 & 2.38\% & [S23]\\
\textbf{Andrea De Lucia} & 1 & 2.38\% & [S09]\\
\textbf{Andrew Meneely} & 1 & 2.38\% & [S27]\\
\textbf{Anna Perini} & 1 & 2.38\% & [S18]\\
\textbf{Antonio Capizzi} & 1 & 2.38\% & [S37]\\
\textbf{Arianne Navarro Lepe} & 1 & 2.38\% & [S33]\\
\textbf{Ashish Sureka} & 1 & 2.38\% & [S16]\\
\textbf{Ayrton Mondragon Mejia} & 1 & 2.38\% & [S33]\\
\textbf{Benjamin C. M. Fung} & 1 & 2.38\% & [S14]\\
\textbf{Bram Adams} & 1 & 2.38\% & [S04]\\
\textbf{Burak Turhan} & 1 & 2.38\% & [S15]\\
\textbf{Byron J. Williams} & 1 & 2.38\% & [S01]\\
\textbf{Chakkrit Tantithamthavorn} & 1 & 2.38\% & [S20]\\
\textbf{Chang Xu} & 1 & 2.38\% & [S10]\\
\textbf{Christoph Treude} & 1 & 2.38\% & [S06]\\
\textbf{Chunyang Chen} & 1 & 2.38\% & [S30]\\
\textbf{Cosmo D'Uva} & 1 & 2.38\% & [S09]\\
\textbf{Daniel Izquierdo-Cortazar} & 1 & 2.38\% & [S26]\\
\textbf{Dario Di Nucci} & 1 & 2.38\% & [S09]\\
\textbf{Davide Fucci} & 1 & 2.38\% & [S15]\\
\textbf{Deheng Ye} & 1 & 2.38\% & [S22]\\
\textbf{Ejaz ul Haq} & 1 & 2.38\% & [S35]\\
\textbf{Eklavya Bhatia} & 1 & 2.38\% & [S35]\\
\textbf{Eman Abdullah AlOmar} & 1 & 2.38\% & [S42]\\
\textbf{Evgeny Bobrov} & 1 & 2.38\% & [S37]\\
\textbf{Fabio Palomba} & 1 & 2.38\% & [S09]\\
\textbf{Ferdian Thung} & 1 & 2.38\% & [S06]\\
\textbf{Filomena Ferrucci} & 1 & 2.38\% & [S09]\\
\textbf{Fitsum Meshesha Kifetew} & 1 & 2.38\% & [S18]\\
\textbf{Garvit Rana} & 1 & 2.38\% & [S35]\\
\textbf{Gede Artha Azriadi Prana} & 1 & 2.38\% & [S06]\\
\textbf{Hareem Sahar} & 1 & 2.38\% & [S41]\\
\textbf{Heng Yin} & 1 & 2.38\% & [S39]\\
\textbf{Hongyu Zhang} & 1 & 2.38\% & [S07]\\
\textbf{Iman Keivanloo} & 1 & 2.38\% & [S05]\\
\textbf{Ismael Solis Moreno} & 1 & 2.38\% & [S33]\\
\textbf{Itzel Morales-Ramirez} & 1 & 2.38\% & [S18]\\
\textbf{Javaid Iqbal Bhat} & 1 & 2.38\% & [S34]\\
\textbf{Jesus M. Gonzalez-Barahona} & 1 & 2.38\% & [S26]\\
\textbf{Jiahuan He} & 1 & 2.38\% & [S31]\\
\textbf{Jian-Ling Sun} & 1 & 2.38\% & [S29]\\
\textbf{Jian Zhang} & 1 & 2.38\% & [S10]\\
\textbf{Jing Jiang} & 1 & 2.38\% & [S31]\\
\textbf{Jorge Luis Victória Barbosa} & 1 & 2.38\% & [S38]\\
\textbf{Jue Wang} & 1 & 2.38\% & [S10]\\
\textbf{Jun Yan} & 1 & 2.38\% & [S10]\\
\textbf{Kaisar Javeed Giri} & 1 & 2.38\% & [S34]\\
\textbf{Karim Ali} & 1 & 2.38\% & [S41]\\
\textbf{Kazi Zakia Sultana} & 1 & 2.38\% & [S01]\\
\textbf{Kleinner Silva Farias de Oliveira} & 1 & 2.38\% & [S38]\\
\textbf{Lars Kurth} & 1 & 2.38\% & [S26]\\
\textbf{Latifa Guerrouj} & 1 & 2.38\% & [S14]\\
\textbf{Leandro Ferreira D'Avila} & 1 & 2.38\% & [S38]\\
\textbf{Li Zhang} & 1 & 2.38\% & [S31]\\
\textbf{Lili Wei} & 1 & 2.38\% & [S10]\\
\textbf{Luis Cruz} & 1 & 2.38\% & [S24]\\
\textbf{Luiz J. P. Araújo} & 1 & 2.38\% & [S37]\\
\textbf{Manuel Mazzara} & 1 & 2.38\% & [S37]\\
\textbf{Megha Mittal} & 1 & 2.38\% & [S16]\\
\textbf{Meiyappan Nagappan} & 1 & 2.38\% & [S28]\\
\textbf{Meng Yan} & 1 & 2.38\% & [S08]\\
\textbf{MingWen} & 1 & 2.38\% & [S07]\\
\textbf{Mohamed Wiem Mkaouer} & 1 & 2.38\% & [S42]\\
\textbf{Muhammad Ahmad} & 1 & 2.38\% & [S37]\\
\textbf{Nachiket Kapre} & 1 & 2.38\% & [S22]\\
\textbf{Nasir Ali} & 1 & 2.38\% & [S12]\\
\textbf{Nelson Sekitoleko} & 1 & 2.38\% & [S26]\\
\textbf{Nuthan Munaiah} & 1 & 2.38\% & [S27]\\
\textbf{Pasquale Salza} & 1 & 2.38\% & [S09]\\
\textbf{Patanamon Thongtanunam} & 1 & 2.38\% & [S32]\\
\textbf{Patricia Ortegon Cano} & 1 & 2.38\% & [S33]\\
\textbf{Pavneet Singh Kochhar} & 1 & 2.38\% & [S31]\\
\textbf{Rahul Katarya} & 1 & 2.38\% & [S35]\\
\textbf{Rahul Krishna} & 1 & 2.38\% & [S40]\\
\textbf{Rodrigo Morales} & 1 & 2.38\% & [S13]\\
\textbf{Rohit Vashisht} & 1 & 2.38\% & [S36]\\
\textbf{Romain Robbes} & 1 & 2.38\% & [S23]\\
\textbf{RongxinWu} & 1 & 2.38\% & [S07]\\
\textbf{Rubén Saborido} & 1 & 2.38\% & [S13]\\
\textbf{Rui Abreu} & 1 & 2.38\% & [S24]\\
\textbf{Salvatore Distefano} & 1 & 2.38\% & [S37]\\
\textbf{Seyyed Ehsan Salamati Taba} & 1 & 2.38\% & [S05]\\
\textbf{Shaohua Wang} & 1 & 2.38\% & [S05]\\
\textbf{Silvana De Gyves Avila} & 1 & 2.38\% & [S33]\\
\textbf{Stuart McIlroy} & 1 & 2.38\% & [S12]\\
\textbf{Sunghee Lee} & 1 & 2.38\% & [S02]\\
\textbf{Syed Afzal Murtaza Rizvi} & 1 & 2.38\% & [S36]\\
\textbf{Tanmay Bhowmik} & 1 & 2.38\% & [S01]\\
\textbf{Thushari Atapattu} & 1 & 2.38\% & [S06]\\
\textbf{Tianyong Wu} & 1 & 2.38\% & [S10]\\
\textbf{Tim Menzies} & 1 & 2.38\% & [S40]\\
\textbf{Tse-Hsun (Peter) Chen} & 1 & 2.38\% & [S19]\\
\textbf{Venera Arnaoudova} & 1 & 2.38\% & [S14]\\
\textbf{Wei Wu} & 1 & 2.38\% & [S04]\\
\textbf{Woo Jin Lee} & 1 & 2.38\% & [S02]\\
\textbf{Xin-Li Yang} & 1 & 2.38\% & [S29]\\
\textbf{Yang Liu} & 1 & 2.38\% & [S30]\\
\textbf{Yepang Liu} & 1 & 2.38\% & [S10]\\
\textbf{Yu QU} & 1 & 2.38\% & [S39]\\
\textbf{Yuan Tian} & 1 & 2.38\% & [S28]\\
\textbf{Yuanrui Fan} & 1 & 2.38\% & [S11]\\
\textbf{Zahid Hussain Wani} & 1 & 2.38\% & [S34]\\
\textbf{Zeinab Kermansaravi} & 1 & 2.38\% & [S14]\\
\textbf{Zhi-Yuan Wan} & 1 & 2.38\% & [S29]\\

\end{longtable}



\onecolumn

\begin{longtable}{p{7cm}ccp{6cm}}
     \caption{Statistics per Institution}
     \label{table:SumInstitutions}
    \\\hline\noalign{\smallskip}
		\textbf{Institution} & \textbf{Freq.} & \textbf{Perc.} & \textbf{Ref.}\\
    \noalign{\smallskip}\hline\noalign{\smallskip}
\endfirsthead

\multicolumn{4}{c}
{ \begin{footnotesize} \tablename\ \thetable{}: continued from previous page \end{footnotesize} } \\


   		\textbf{Institution} & \textbf{Freq.} & \textbf{Perc.} & \textbf{Ref.}\\

\noalign{\smallskip}\hline\noalign{\smallskip}
\endhead

\hline \multicolumn{4}{r}{{Continued on next page}} \\ 
\noalign{\smallskip}\hline\noalign{\smallskip}
\endfoot

\hline
\endlastfoot

   \textbf{Queen's University} & 11 & 26.19\% & [S03], [S05], [S08], [S12], [S17], [S19], [S20], [S21], [S25], [S28], [S32]\\
\textbf{Singapore Management University} & 7 & 16.67\% & [S06], [S08], [S11], [S24], [S28], [S29], [S31]\\
\textbf{Concordia University} & 4 & 9.52\% & [S03], [S19], [S25], [S32]\\
\textbf{Zhejiang University} & 4 & 9.52\% & [S08], [S11], [S29], [S31]\\
\textbf{École Polytechnique de Montréal} & 3 & 7.14\% & [S04], [S13], [S14]\\
\textbf{Monash University} & 3 & 7.14\% & [S08], [S11], [S30]\\
\textbf{Rochester Institute of Technology} & 3 & 7.14\% & [S27], [S28], [S42]\\
\textbf{Hong Kong University of Science and Technology} & 2 & 4.76\% & [S07], [S10]\\
\textbf{Nanyang Technological University} & 2 & 4.76\% & [S22], [S30]\\
\textbf{University of Adelaide} & 2 & 4.76\% & [S06], [S20]\\
\textbf{University of Zurich} & 2 & 4.76\% & [S09], [S23]\\
\textbf{Australian National University} & 1 & 2.38\% & [S30]\\
\textbf{Beihang University} & 1 & 2.38\% & [S31]\\
\textbf{Bitergia} & 1 & 2.38\% & [S26]\\
\textbf{Citrix} & 1 & 2.38\% & [S26]\\
\textbf{Columbia University} & 1 & 2.38\% & [S40]\\
\textbf{Delft University of Technology} & 1 & 2.38\% & [S23]\\
\textbf{Delhi Technological University} & 1 & 2.38\% & [S35]\\
\textbf{École de Technologie Supérieure} & 1 & 2.38\% & [S14]\\
\textbf{ETS Montreal, University of Quebec} & 1 & 2.38\% & [S42]\\
\textbf{Fondazione Bruno Kessler} & 1 & 2.38\% & [S18]\\
\textbf{Free University of Bozen-Bolzano} & 1 & 2.38\% & [S23]\\
\textbf{IBM} & 1 & 2.38\% & [S33]\\
\textbf{Indraprastha Institute of Information Technology} & 1 & 2.38\% & [S16]\\
\textbf{INESC ID} & 1 & 2.38\% & [S24]\\
\textbf{INFOTEC} & 1 & 2.38\% & [S18]\\
\textbf{Innopolis University} & 1 & 2.38\% & [S37]\\
\textbf{Islamic University of Science and Technology} & 1 & 2.38\% & [S34]\\
\textbf{Jamia Millia Islamia} & 1 & 2.38\% & [S36]\\
\textbf{Kyungpook National University} & 1 & 2.38\% & [S02]\\
\textbf{McGill University} & 1 & 2.38\% & [S14]\\
\textbf{Mississippi State University} & 1 & 2.38\% & [S01]\\
\textbf{Nanjing University} & 1 & 2.38\% & [S10]\\
\textbf{NC State University} & 1 & 2.38\% & [S40]\\
\textbf{Southern University of Science and Technology} & 1 & 2.38\% & [S10]\\
\textbf{Universidad Rey Juan Carlos} & 1 & 2.38\% & [S26]\\
\textbf{Universitá della Svizzera Italiana} & 1 & 2.38\% & [S09]\\
\textbf{University of Alberta} & 1 & 2.38\% & [S41]\\
\textbf{University of California} & 1 & 2.38\% & [S39]\\
\textbf{University of Chinese Academy of Sciences} & 1 & 2.38\% & [S10]\\
\textbf{University of Lisbon} & 1 & 2.38\% & [S24]\\
\textbf{University of Melbourne} & 1 & 2.38\% & [S32]\\
\textbf{University of Messina} & 1 & 2.38\% & [S37]\\
\textbf{University of Newcastle} & 1 & 2.38\% & [S07]\\
\textbf{University of Oulu} & 1 & 2.38\% & [S15]\\
\textbf{University of Salerno} & 1 & 2.38\% & [S09]\\
\textbf{University of Vale do Rio dos Sinos} & 1 & 2.38\% & [S38]\\
\textbf{University of Waterloo} & 1 & 2.38\% & [S12]\\
\textbf{Vrije Universiteit Brussel} & 1 & 2.38\% & [S09]\\
\textbf{Washington State University} & 1 & 2.38\% & [S14]\\


\end{longtable}




\onecolumn

\begin{longtable}{p{3cm}ccp{9cm}}
    \caption{Statistics per Continent and Country}
\label{table:SumCountries}
    \\\hline\noalign{\smallskip}
		\textbf{} & \textbf{Freq.} & \textbf{Perc.} & \textbf{Ref.}\\
    \noalign{\smallskip}\hline\noalign{\smallskip}
\endfirsthead

\multicolumn{4}{c}
{ \begin{footnotesize} \tablename\ \thetable{}: continued from previous page \end{footnotesize} } \\

\hline\noalign{\smallskip}

   		\textbf{Name} & \textbf{Freq.} & \textbf{Perc.} & \textbf{Ref.}\\

\noalign{\smallskip}\hline\noalign{\smallskip}
\endhead

\hline \multicolumn{4}{r}{{Continued on next page}} \\ 
\noalign{\smallskip}\hline\noalign{\smallskip}
\endfoot

\hline
\endlastfoot
 
 \multicolumn{4}{l}{\cellcolor{gray!10}\textbf{Continent}}\\[0.1cm]
 \textbf{North America} & 23 & 54.76\% & [S01], [S03], [S04], [S05], [S08], [S12], [S13], [S14], [S17], [S18], [S19], [S20], [S21], [S25], [S26], [S27], [S28], [S32], [S33], [S39], [S40], [S41], [S42]\\
\textbf{Asia} & 17 & 40.48\% & [S02], [S06], [S07], [S08], [S10], [S11], [S16], [S22], [S24], [S28], [S29], [S30], [S31], [S34], [S35], [S36], [S37]\\
\textbf{Europe} & 7 & 16.67\% & [S09], [S15], [S18], [S23], [S24], [S26], [S37]\\
\textbf{Oceania} & 7 & 16.67\% & [S06], [S07], [S08], [S11], [S20], [S30], [S32]\\
\textbf{South America} & 1 & 2.38\% & [S38]\\
\\\\
 
 \multicolumn{4}{l}{\cellcolor{gray!10}\textbf{Country}}\\[0.1cm]
 \textbf{Canada} & 16 & 38.1\% & [S03], [S04], [S05], [S08], [S12], [S13], [S14], [S17], [S19], [S20], [S21], [S25], [S28], [S32], [S41], [S42]\\
\textbf{Singapore} & 9 & 21.43\% & [S06], [S08], [S11], [S22], [S24], [S28], [S29], [S30], [S31]\\
\textbf{USA} & 8 & 19.05\% & [S01], [S14], [S26], [S27], [S28], [S39], [S40], [S42]\\
\textbf{Australia} & 7 & 16.67\% & [S06], [S07], [S08], [S11], [S20], [S30], [S32]\\
\textbf{China} & 6 & 14.29\% & [S07], [S08], [S10], [S11], [S29], [S31]\\
\textbf{India} & 4 & 9.52\% & [S16], [S34], [S35], [S36]\\
\textbf{Italy} & 4 & 9.52\% & [S09], [S18], [S23], [S37]\\
\textbf{Mexico} & 2 & 4.76\% & [S18], [S33]\\
\textbf{Switzerland} & 2 & 4.76\% & [S09], [S23]\\
\textbf{Belgium} & 1 & 2.38\% & [S09]\\
\textbf{Brazil} & 1 & 2.38\% & [S38]\\
\textbf{Finland} & 1 & 2.38\% & [S15]\\
\textbf{Portugal} & 1 & 2.38\% & [S24]\\
\textbf{Republic of Korea} & 1 & 2.38\% & [S02]\\
\textbf{Russian} & 1 & 2.38\% & [S37]\\
\textbf{Spain} & 1 & 2.38\% & [S26]\\
\textbf{The Netherlands} & 1 & 2.38\% & [S23]\\

\end{longtable}

\begin{table}[!htbp]
\caption{Data Sources Findings (Frequency $>$ 1)}
\label{table:SumSources}
\centering
\begin{tabular}{p{5cm}ccp{8cm}}
	\hline\noalign{\smallskip}
	\textbf{Data Sources} & \textbf{Freq.} & \textbf{Perc.} & \textbf{Ref.}\\
	\noalign{\smallskip}\hline\noalign{\smallskip}
 
 \textbf{GitHub Repositories} & 10 & 23.81\% & [S02], [S06], [S09], [S10], [S13], [S23], [S24], [S31], [S37], [S42]\\[0.1cm]
\textbf{Google Play Store} & 7 & 16.67\% & [S03], [S05], [S10], [S12], [S20], [S24], [S28]\\[0.1cm]
\textbf{Git Repositories} & 6 & 14.29\% & [S08], [S14], [S16], [S19], [S26], [S32]\\[0.1cm]
\textbf{BugZilla} & 5 & 11.9\% & [S07], [S14], [S16], [S17], [S21]\\[0.1cm]
\textbf{F-Droid Repository} & 5 & 11.9\% & [S02], [S03], [S09], [S10], [S24]\\[0.1cm]
\textbf{Promise Repositories} & 4 & 9.52\% & [S35], [S39], [S40], [S41]\\[0.1cm]
\textbf{Online Survey} & 3 & 7.14\% & [S15], [S23], [S24]\\[0.1cm]
\textbf{StackOverflow} & 3 & 7.14\% & [S22], [S29], [S30]\\[0.1cm]
\textbf{JAVA} & 2 & 4.76\% & [S35], [S38]\\[0.1cm]
\textbf{Maven Repositories} & 2 & 4.76\% & [S04], [S09]\\[0.1cm]
\textbf{SVN Repositories} & 2 & 4.76\% & [S09], [S14]\\[0.1cm]
\textbf{Unknown} & 2 & 4.76\% & [S34], [S36]\\[0.1cm]
\textbf{Android Issue Tracker} & 1 & 2.38\% & [S27]\\[0.1cm]
\textbf{Apache OpenOffice Issue Tracking System} & 1 & 2.38\% & [S18]\\[0.1cm]
\textbf{Apache Tomcat Archive} & 1 & 2.38\% & [S01]\\[0.1cm]
\textbf{BinTray} & 1 & 2.38\% & [S09]\\[0.1cm]
\textbf{Cassandra} & 1 & 2.38\% & [S33]\\[0.1cm]
\textbf{Chrome Releases Blog} & 1 & 2.38\% & [S27]\\[0.1cm]
\textbf{Chromium Issue Tracker} & 1 & 2.38\% & [S27]\\[0.1cm]
\textbf{CodeClimate} & 1 & 2.38\% & [S37]\\[0.1cm]
\textbf{Docker} & 1 & 2.38\% & [S37]\\[0.1cm]
\textbf{Eclise API} & 1 & 2.38\% & [S38]\\[0.1cm]
\textbf{Exception Reports} & 1 & 2.38\% & [S07]\\[0.1cm]
\textbf{Gerrit} & 1 & 2.38\% & [S11]\\[0.1cm]
\textbf{Google} & 1 & 2.38\% & [S09]\\[0.1cm]
\textbf{Google Forms} & 1 & 2.38\% & [S13]\\[0.1cm]
\textbf{HackerOne Bug Bounty Platform} & 1 & 2.38\% & [S27]\\[0.1cm]
\textbf{JCenter} & 1 & 2.38\% & [S09]\\[0.1cm]
\textbf{Jenkins} & 1 & 2.38\% & [S37]\\[0.1cm]
\textbf{JIRA} & 1 & 2.38\% & [S14]\\[0.1cm]
\textbf{Lab Computers} & 1 & 2.38\% & [S15]\\[0.1cm]
\textbf{Mailing List} & 1 & 2.38\% & [S26]\\[0.1cm]
\textbf{Mercurial Repositories} & 1 & 2.38\% & [S16]\\[0.1cm]
\textbf{MongoDB} & 1 & 2.38\% & [S33]\\[0.1cm]
\textbf{Mylyn} & 1 & 2.38\% & [S38]\\[0.1cm]
\textbf{NetBeans Source Code Repository} & 1 & 2.38\% & [S07]\\[0.1cm]
\textbf{Node} & 1 & 2.38\% & [S37]\\[0.1cm]
\textbf{Python} & 1 & 2.38\% & [S33]\\[0.1cm]
\textbf{SEACRAFT Repositories} & 1 & 2.38\% & [S35]\\[0.1cm]
\textbf{SecuriBench Archive} & 1 & 2.38\% & [S01]\\[0.1cm]
\textbf{SEnerCON Feedback Gathering System} & 1 & 2.38\% & [S18]\\[0.1cm]
\textbf{Slack} & 1 & 2.38\% & [S37]\\[0.1cm]
\textbf{Spark} & 1 & 2.38\% & [S33]\\[0.1cm]
\textbf{Team Wiki (BitBucket)} & 1 & 2.38\% & [S16]\\[0.1cm]
\textbf{Version Control Repositories} & 1 & 2.38\% & [S25]\\[0.1cm]
\textbf{Vulnerability Reports} & 1 & 2.38\% & [S01]\\[0.1cm]
\hline

\end{tabular}
\end{table}

%
%
%
%
%
%
%
%
%

\begin{landscape}

\plot{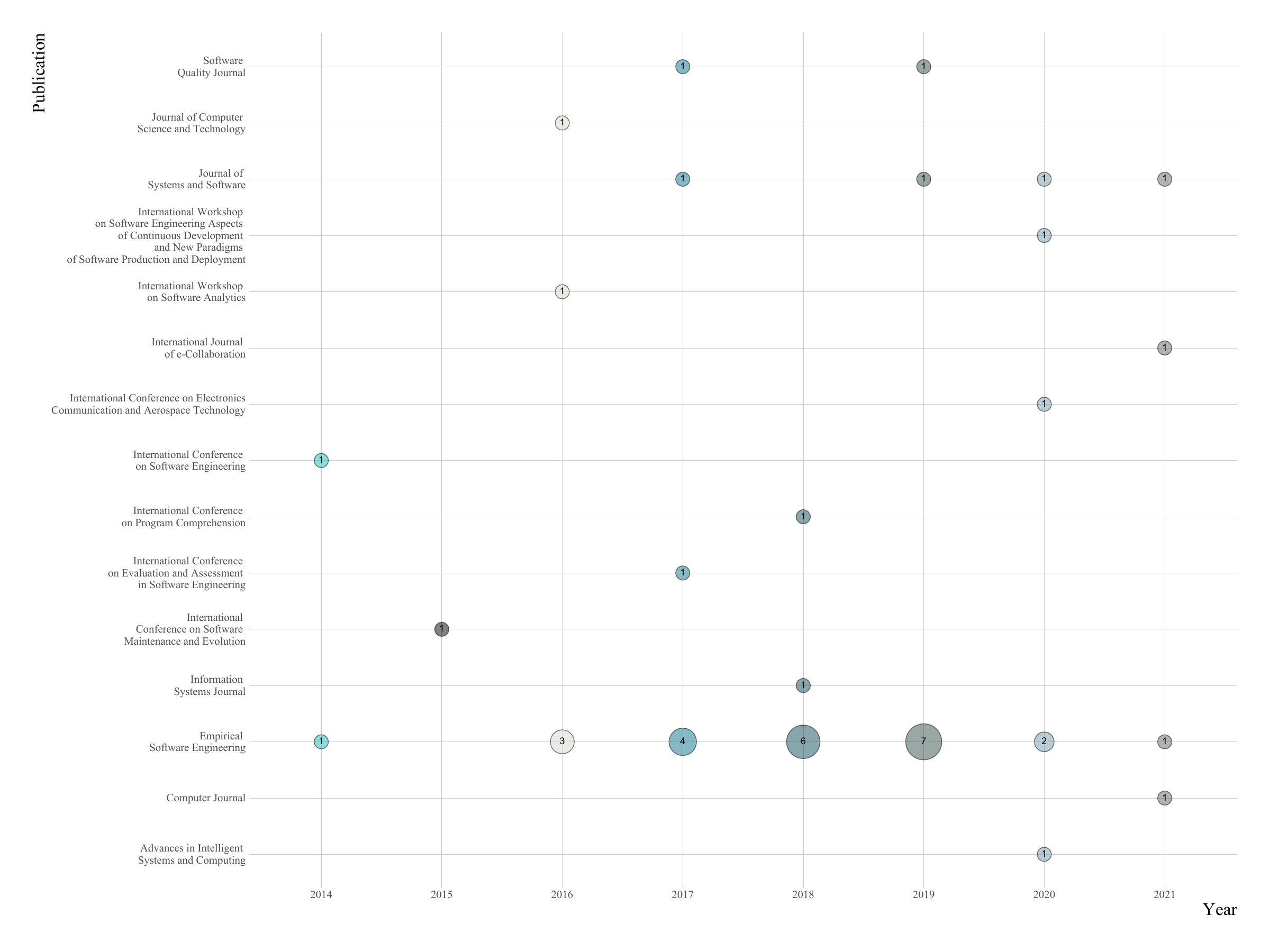}[Frequencies of studies per Publisher over the Years][22][H][trim=1cm 1cm 1cm 1cm]

\plot{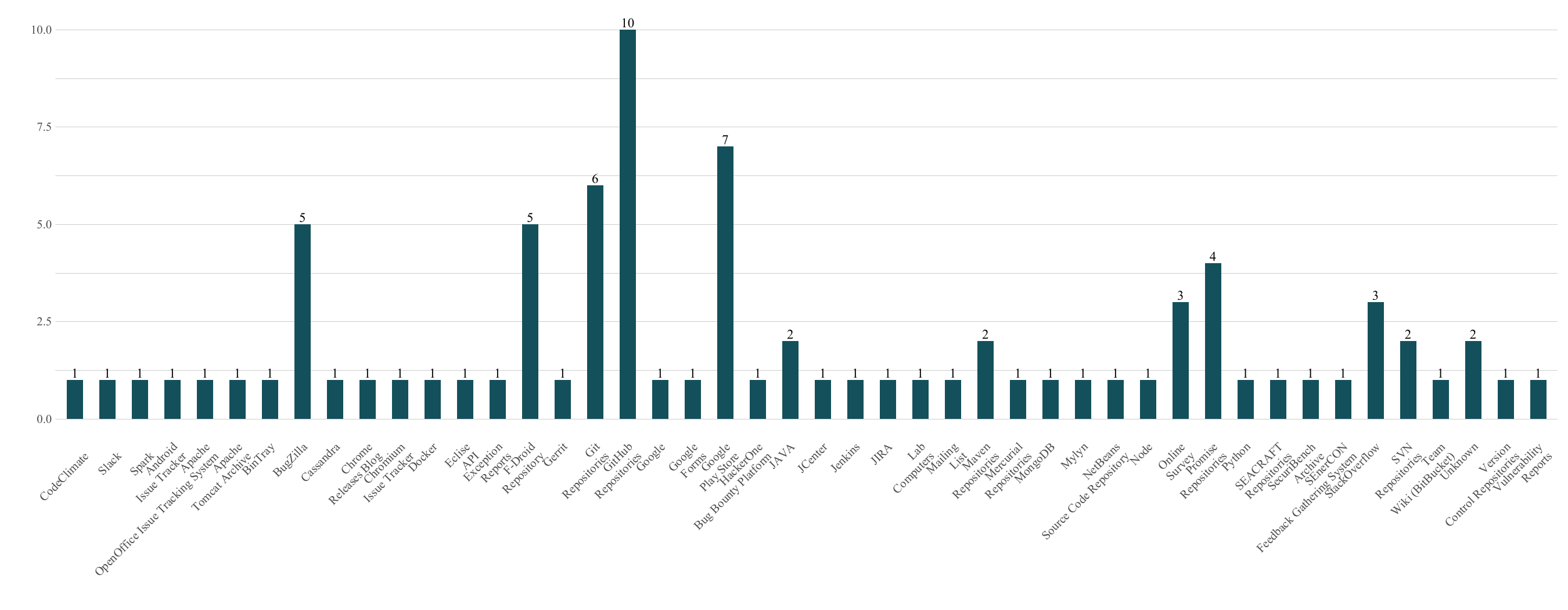}[Frequencies of studies for Data Sources][22][H][trim=0cm 0.9cm 0cm 0.4cm]

\plot{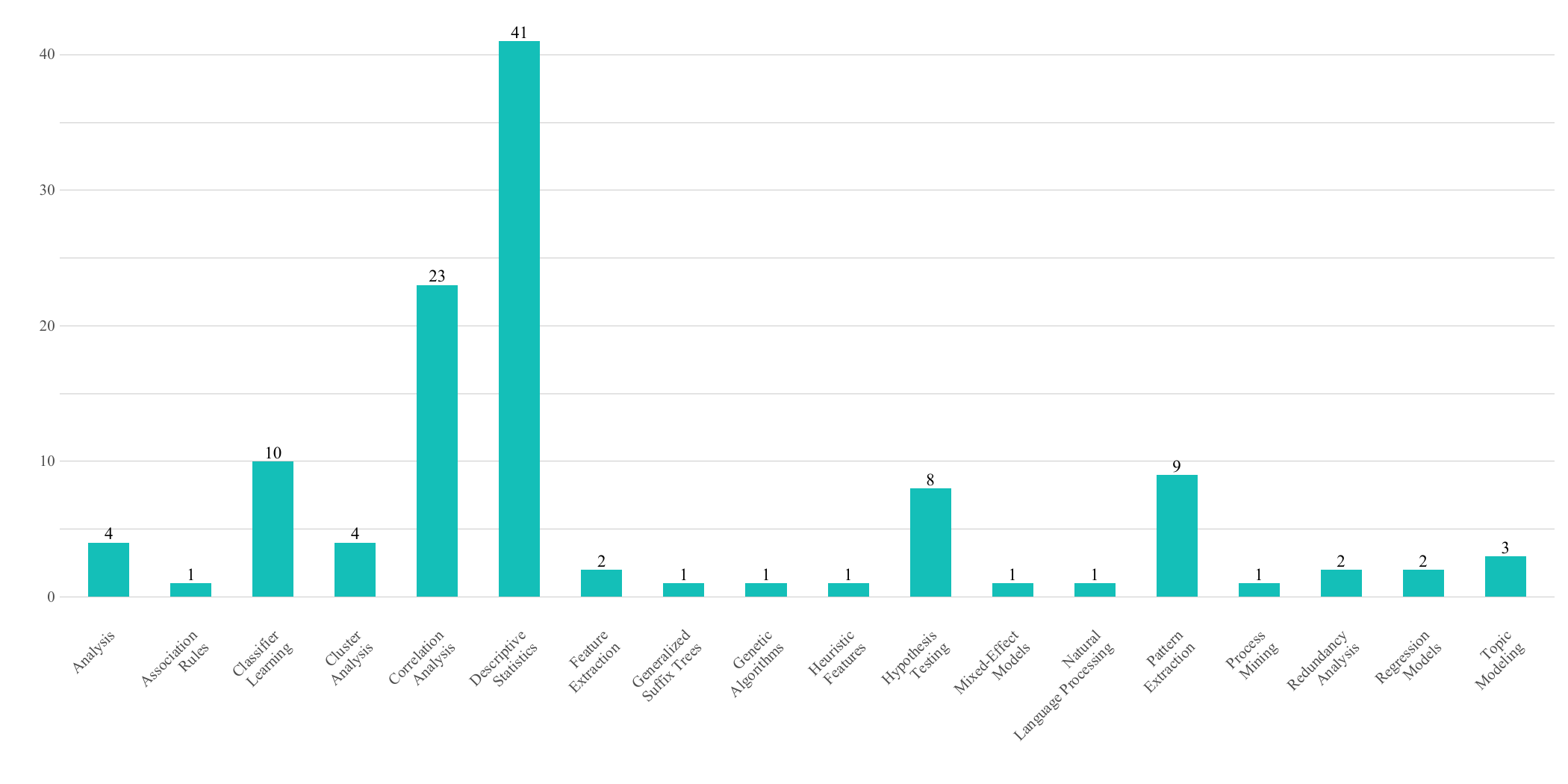}[Frequencies of studies for Mining Methods][22][H][trim=0cm 0.9cm 0cm 0.6cm]

\end{landscape}

\plot{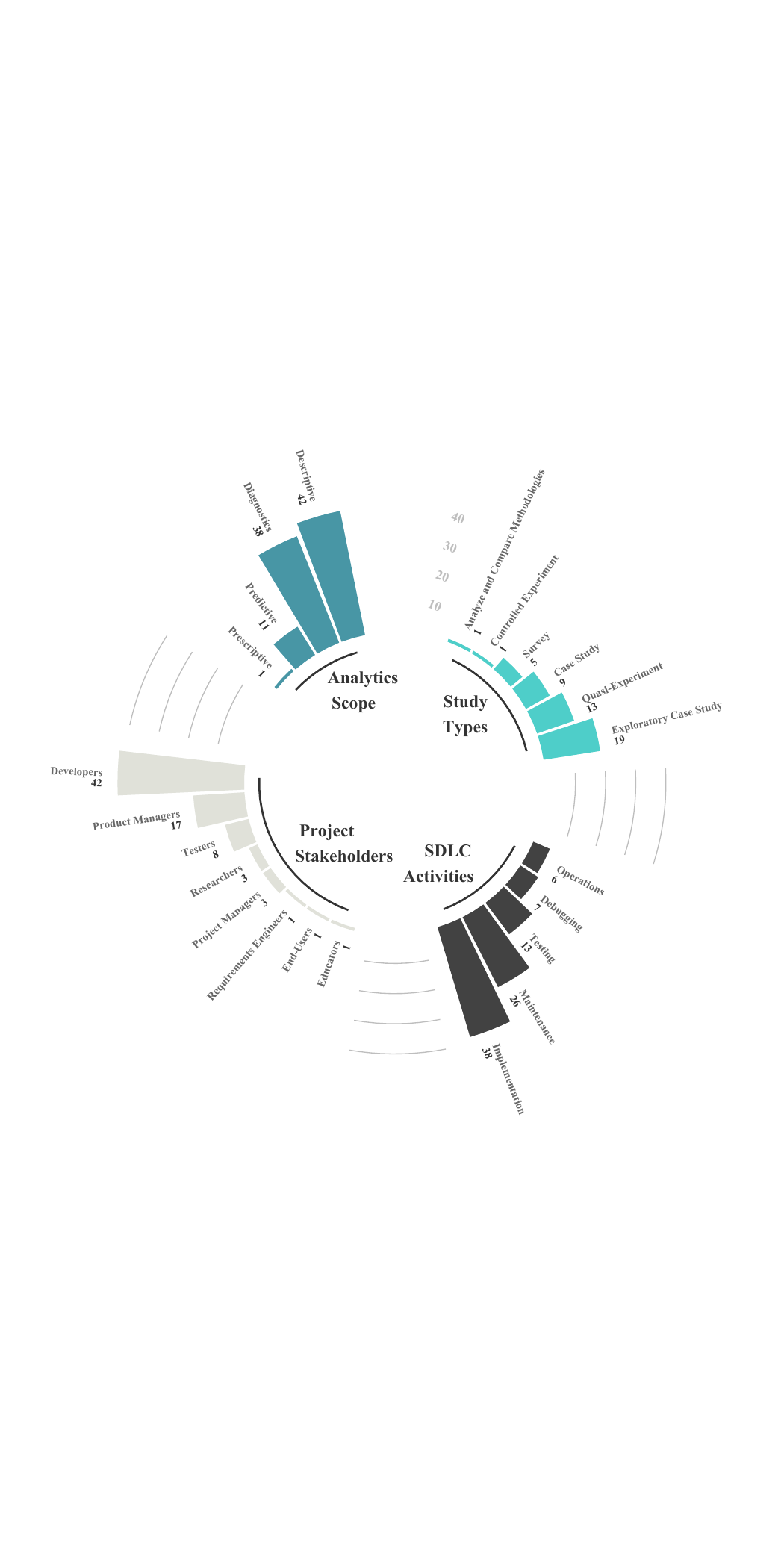}[Frequencies of studies combining multiple RQs in the SLR][16][h!][trim=1.1cm 10cm 1.3cm 10cm]


\newpage
\normalsize

\section{Studies Appraisal}
\label{sec:AppendixC}

The following acronyms were used for SLR results interpretation:

\begin{itemize}
    \item \textbf{Study Type} 
     \subitem{\textbf{ACM}-Analyze and Compare Methodologies, \textbf{CS}-Case Study, \textbf{CE}-Controlled Experiment
      \subitem{\textbf{ECS}-Exploratory Case Study, \textbf{QE}-Quasi-Experiment, \textbf{S}-Survey}}
    
    \item \textbf{SDLCActivities}  
    \subitem{\textbf{D}-Debugging, \textbf{I}-Implementation, \textbf{M}-Maintenance, \textbf{O}-Operations, \textbf{T}-Testing}
    
    \item \textbf{Project Stakeholders}
    
    \subitem{\textbf{D}-Developers, \textbf{E}-Educators, \textbf{EU}-End-Users, \textbf{T}-Testers, \textbf{PM}-Product Managers} 
    \subitem{\textbf{PjM}-Project Managers, \textbf{R}-Researchers,  \textbf{RE}-Requirements Engineers}
    
    \item \textbf{Analytics Scope} 
    \subitem{\textbf{Des}-Descriptive Analytics, \textbf{Dia}-Diagnostics Analytics}
    \subitem{\textbf{Pred}-Predictive Analytics, \textbf{Pres}-Prescriptive Analytics}\\
\end{itemize}

The following taxonomy was used to assess the SDLC contributions:

\NumTabs{5}
\begin{itemize}
     \item \textbf{The benefit is:} 
	 \subitem{\textbf{Absent (0)} \tabto{11em} \zeropct  \hspace{5mm} Not addressed}
     \subitem{\textbf{Weak (0.25)} \tabto{11em} \quarterpct \hspace{5mm} Implicitly addressed}
     \subitem{\textbf{Moderate (0.5)} \tabto{11em} \halfpct \hspace{5mm} Explicitly addressed (not detailed)}
     \subitem{\textbf{Strong (0.75)} \tabto{11em} \threequarterpct \hspace{5mm} Explained with details and implications}
     \subitem{\textbf{Complete (1)} \tabto{11em} \fullpct \hspace{5mm} Fully explained, validated and replicable}
\end{itemize}

\begin{landscape}

\scriptsize
\begin{longtable}{ccp{2cm}cccp{2.7cm}ccccccc}
    \caption{Systematic Literature Review Results.}
	\label{tab:slr:results}
    \\\hline\noalign{\smallskip}
     &
	\multicolumn{1}{c}{\textbf{\shortstack[c]{Study \\Type}}} & 
    \multicolumn{1}{c}{\textbf{\shortstack[c]{Data \\Sources}}} &
    \multicolumn{1}{c}{\textbf{\shortstack[c]{Process \\Perspective}}} &
    \multicolumn{1}{c}{\textbf{\shortstack[c]{SDLC \\Activities}}} &
	\multicolumn{1}{c}{\textbf{\shortstack[c]{Project \\Stakeholders}}} &
	\multicolumn{1}{c}{\textbf{\shortstack[c]{Mining \\Methods}}} &	
    \multicolumn{1}{c}{\textbf{\shortstack[c]{Analytics \\Scope}}} &
    \multicolumn{5}{c}{\textbf{\shortstack[c]{Contributions \\to SDLC}}} \\
	\multicolumn{1}{c}{\textbf{Study}} &    &    &    &    &    &    &    &
    \cellcolor{black!10}\rotatebox{90}{\shortstack[c]{Technical \\Debt}} & 
    \cellcolor{black!10}\rotatebox{90}{\shortstack[c]{Time \\Management}} & 
    \cellcolor{black!10}\rotatebox{90}{\shortstack[c]{Costs \\Control}} & 
    \cellcolor{black!10}\rotatebox{90}{\shortstack[c]{Risks \\Assessment}} & 
    \cellcolor{black!10}\rotatebox{90}{\shortstack[c]{Security \\Analysis}} \\
    
\noalign{\smallskip}\hline\noalign{\smallskip}
\endfirsthead

\multicolumn{13}{c}
{{\tablename\ \thetable{}: continued from previous page}} \\
    \hline\noalign{\smallskip}
     &
	\multicolumn{1}{c}{\textbf{\shortstack[c]{Study \\Type}}} & 
    \multicolumn{1}{c}{\textbf{\shortstack[c]{Data \\Sources}}} &
    \multicolumn{1}{c}{\textbf{\shortstack[c]{Process \\Perspective}}} &
    \multicolumn{1}{c}{\textbf{\shortstack[c]{SDLC \\Activities}}} &
	\multicolumn{1}{c}{\textbf{\shortstack[c]{Project \\Stakeholders}}} &
	\multicolumn{1}{c}{\textbf{\shortstack[c]{Mining \\Methods}}} &	
    \multicolumn{1}{c}{\textbf{\shortstack[c]{Analytics \\Scope}}} &
    \multicolumn{5}{c}{\textbf{\shortstack[c]{Contributions \\to SDLC}}} \\
    \multicolumn{1}{c}{\textbf{Study}} &    &    &    &    &    &    &    &
    \cellcolor{black!10}\rotatebox{90}{\shortstack[c]{Technical \\Debt}} & 
    \cellcolor{black!10}\rotatebox{90}{\shortstack[c]{Time \\Management}} & 
    \cellcolor{black!10}\rotatebox{90}{\shortstack[c]{Costs \\Control}} & 
    \cellcolor{black!10}\rotatebox{90}{\shortstack[c]{Risks \\Assessment}} & 
    \cellcolor{black!10}\rotatebox{90}{\shortstack[c]{Security \\Analysis}} \\
    
\noalign{\smallskip}\hline\noalign{\smallskip}
\endhead

\hline \multicolumn{13}{r}{{Continued on next page}} \\ 
\noalign{\smallskip}\hline\noalign{\smallskip}
\endfoot

\hline
\endlastfoot

\end{longtable}

\normalsize

\twocolumn

\end{landscape}

\end{document}